%% file: draft.tex
\documentclass{article}
\usepackage{hyperref}
\usepackage{graphicx}
\usepackage{listings}
\usepackage{amssymb}
\usepackage{bbm}
\usepackage{rotating}
\usepackage{amsmath}
\usepackage{caption}
\usepackage[position=top]{subcaption}
\begin{document}
\title{Measuring the Demand Effects of Formal and Informal Communication : Evidence from Online Markets for Illicit Drugs}
\author{Luis Armona \\ \href{mailto:larmona@stanford.edu}{larmona@stanford.edu} }
\maketitle

\begin{abstract}
I present evidence that communication between marketplace participants is an important influence on market demand. I find that consumer demand is approximately equally influenced by communication on both formal and informal networks- namely, product reviews and community forums. In addition, I find empirical evidence of a vendor's ability to commit to disclosure dampening the effect of communication on demand. I also find that product demand is more responsive to average customer sentiment as the number of messages grows, as may be expected in a Bayesian updating framework.
\end{abstract}

\section{Introduction}

Product information is essential for markets to function well. Without accurate knowledge of what they are buying, consumers cannot be expected to maximize personal welfare over the set of product choices. Of course, product characteristics that affect a consumer's ability to optimize in a marketplace go beyond knowledge of its individual components. In the absence of strong institutions, consumer's might greatly value a seller's reputation, or be concerned that a seller might cheat them, particularly if they are unable to observe product quality at time of purchase.
Of critical importance in this setting would be the avenues through which consumers can collect information about the product space they seek to purchase from.

In this paper, I examine a market with exactly these characteristics, and attempt to measure the relative importance of differential modes of information transmission in relation to consumer demand. Specifically, I study the demand for illegal products on online ``Darknet'' markets, which lack credible enforcement of laws and so must rely on incentives from vendor reputation and consumers being appropriately informed to function well,. Because of this and the concentrated set of locations where consumers can gather information about products, I am able to plausibly estimate the impact of product sentiment on future market demand, by controlling for all feasible supply-side decisions that may impact consumer demand. By ``sentiment'', I mean whether a vendor is being talked about in a positive or negative way, as measured by the word structure in the message. I find that both informal and formal communication (as measured by forum post and product review sentiment, respectively) have a large effect on consumer demand, and that these effects are comparable in magnitude. I find that these effects scale with sample size of the information set , but little other evidence for heterogeneity in the effects of product sentiment from varying the information sources.

My marketplace is the illegal online market for illicit products, such as guns, narcotics, and fraud services, commonly known as the ``Darknet'' of the Internet. Due to the illegal nature of all transactions being performed\footnote{even if the product itself is legal in a given country, no government taxes are ever paid by the vendors, so they are always illegal.}, this marketplace is highly anonymous. Vendors and sellers alike are potentially culpable in the distribution of these goods, which in developed countries such as the United States, can result in up to 15 years of imprisonment \cite{csa}. In addition to the clear personal incentives one might have to remain anonymous, anonymity is enforced by requiring users to connect to these online marketplaces via the TOR protocol. TOR is a specific type of Internet routing service that scrambles ones personal data packets  by interchanging them with the data packets of other current active TOR users \cite{dingledine2004tor}. The net result is that an observer, whether a government or a private organization, is unable to tell what IP address a traced data request is coming from, thereby offering an additional layer of security for both customers and vendors. On top of this, all transactions on the marketplace are done in Bitcoins, a popular cryptocurrency notorious for its ability to anonymize transactions.

\begin{figure}
\centering
\caption{A Typical Agora Market Listing}
\label{fig:samplelisting}
\includegraphics[scale=.4]{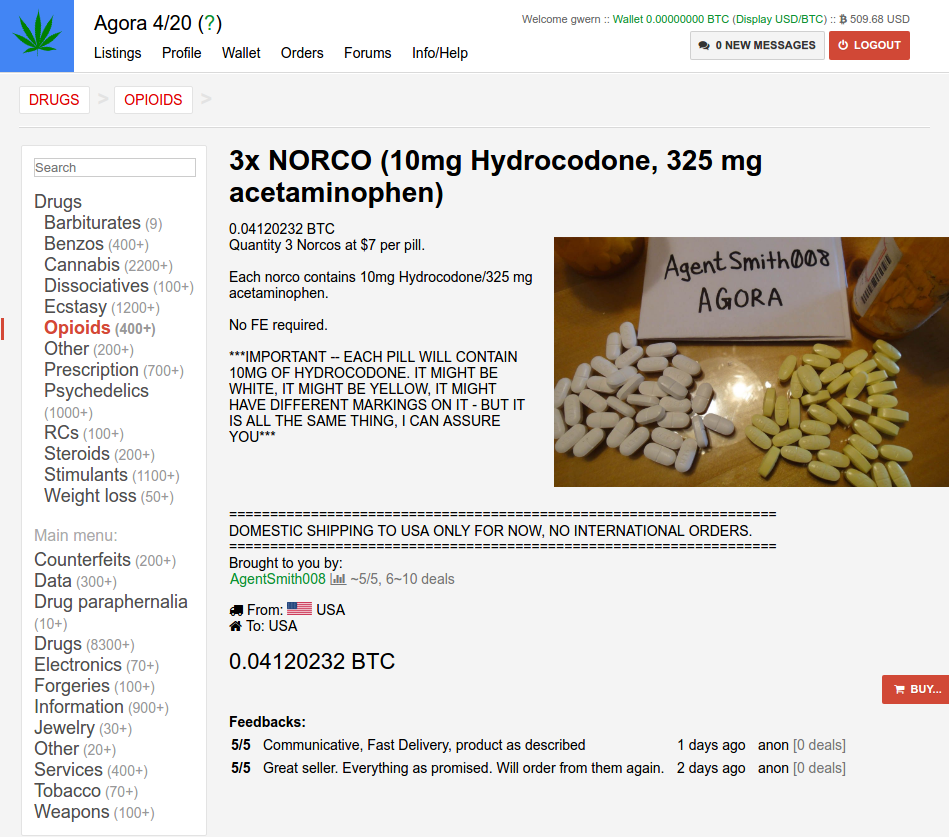}
\end{figure}

Besides the strong protections of anonymity used by the market, Darknet marketplaces operates like virtually any other e-commerce website. Listings are organized into relevant product categories (LSD, Opiods, Firearms, etc.), and customers can use a search engine to locate any specific product they wish to consume or vendor they wish to buy from. On an actual product page, buyers can view a provided description and image of the product from the seller, in addition to the full history of customer reviews from previous buyers, which each have a 0 to 5 star rating, a brief text description provided by previous buyer about their purchase experience and/or the quality of the product, and how long ago the review was made. In addition, the product page has summary statistics on the vendor, such as average rating across all products, total volume of transactions, where the buyer ships to/from in the world, along with a link to a vendor's profile page where one can view past reviews of all previously bought products. Figure \ref{fig:samplelisting} provides an example of a typical listing on the Darknet marketplace I use as my universe for this paper, Agora.

 One rather unique feature of these Darknet markets is that, in order for customers to finalize the transaction, buyers are required to leave a review once they have received the product. In order to ensure the market functions while guaranteeing agent anonymity, Darknet markets use an escrow system, where the marketplace acts as clearinghouse and transfers the Bitcoins from the buyer to the vendor only after the consumer verifies that they have received the product they ordered and leaves their review. The reasons for this are threefold:
 \begin{enumerate}
 \item If every customer is required to leave a review, then the set of reviews are less likely to be biased towards customers with ``extreme'' experiences, either good or bad, and so the reviews might be more informative. Thus, they provide incentives for suppliers to be reliable, and for customers to go ahead with a purchase if they are initially uncertain about the product.
 \item Introducing a third party intermediary further anonymizes the payment process.
 \item If a customer receives an item and has legitimate evidence of the product not matching its advertised description, they may submit a complaint to the marketplace's moderators, and, if found to be true, the complaint will result in the vendor not being paid and possibly banned from the website. This makes fraud costly and further incentivizes vendors to be honest and avoid scamming customers.
 \end{enumerate}
However, a somewhat unfortunate consequence of this well meaning rule is that many vendors explicitly require customers to ``Finalize Early'' when receiving products. This practice involves customers finalizing their transactions with the market clearinghouse in advance of the vendor actually shipping the product, so that the vendor does not have to wait for delivery to receive the payment, a process that can sometimes take weeks. A customer may later edit their review appropriately to match their actual experience, but it is common for a reviewer to mandate that those who finalize early also give a 5-star review; otherwise they will not fulfill the transaction. Part of what originally motivated this paper's investigation into text sentiment was the common practice in Darknet markets of mandating 5-star reviews on finalize early orders, but little done to regulate the text accompanying the review (possibly because it is not an input into a vendor's overall rating, or possibly because it is much more difficult to filter text). Perhaps because of the anarcho-libertarian ideology that surrounded the creation of these Darknet marketplaces, this practice was not explicitly banned by moderators, since advertisers are transparent about this requirement if you wish to purchase from them. Instead, moderators built an explicit ``No Finalize Early'' flag placed on \textit{all} listings in search results so that consumers could easily filter out product offerings that may require finalize early. Misclassifying one's product listing as no finalize early can result in fines to the vendor and in some cases banning the vendor altogether from the market.

One can imagine that all of this taken together implies a huge premium on a vendor's reputation. Despite the protections put in place, this market should still gravitate towards a set of well-established vendors who are well known to serve customers honestly. For this reason, changes in attributes of a vendor reputation (such as community attitudes towards them) should have more pronounced effects in this market, leading to more precise estimations of their effects.


There is a limited literature on Darknet markets due to both their recent inception and illicit nature. Most of the work is done from a criminology perspective, explaining the difference in structures between Darknet markets and traditional illegal drug rings or cartels. One of the first studies of Darknet markets by Christin \cite{christin2013traveling} provides a comprehensive measurement analysis of the Darknet market during its initial stages when it consisted of one website. Demant et al. \cite{demant2016personal} is one of the first papers to examine the Darknet from a social science perspective. They investigate whether consumers on the Darknet are redistributers or direct consumers, and find suggestive evidence that consumers on the Darknet resemble direct-to-consumer sellers, a step below in the drug supply chain.

There is some relevant work that has been done on the importance of communication in online settings. Lewis \cite{lewis2006asymmetric} examines the effects of differential vendor communication or advertisement on sales in eBay's online platform for buying and selling used cars. Luca \& Georgios \cite{luca2016fake} study the incentives of vendors listed on Yelp to procure review fraud in order to boost their own ratings on the website.

\section{Consumer Model}\label{consumermodel}

The market I study is one in which consumers choose to purchase different types of good (mostly different types of drugs).
I assume that demand for product $j$ in time interval $t$ is given by a Poisson distribution:
\[Pr(y_{jt} \textrm{ units purchased} | X = \frac{e^{y\beta X} e^{-e^{\beta X}}}{y_{jt}!} \]
More importantly, the result of this functional form is that our expectation has a straightforward exponential form:
\[E[y_{jt} | X_{j,t},\mu_j] =e^{X_{j,t}\beta + \mu_j }\]
Where $X_{j,t}$ is a vector of time and product varying characteristics, and $\mu_j$ is a product-specific effect on average product demand. More plainly, I model my covariates of interest, $X_{j,t}$, as having a multiplicative $e^\beta$ effect on (mean) product demand for every 1-unit change.

While it is well known that product demand is unobservable, since prices and quantities are results of both demand and supply schedules clearing, I argue in this paper that, due to the unique properties of the market I study, a researcher can observe virtually everything suppliers choose here that are observable to customers. And thus, a researcher can plausibly control for \textit{all} supply-side decisions a seller takes that impact a consumer's choice. Thus, assuming all supply decisions are perfectly controlled for, remaining variation in sales can be explained as a combination of demand-side variation and noise. From here, I take relevant covariates not determined by the supplier, and determine their effect on consumer demand.

My demand-side covariates will mostly examine the effects of peer communication on a vendor's reputation and hence the demand for their products. I examine the impact of two channels of reputation formation: sentiment or satisfaction displayed by past customers on a product's review page, and sentiment concerning a vendor on the designated market forum, which was specifically established so customers could communicate to each other about events or products within the market. I designate these two channels as ``formal'' and ``informal'' communication, since one exists in the formal marketplace setting, where consumers are asked to explicitly review a product, while the other exists in a social setting where users are free to spontaneously discuss whatever they wish. This may sometimes include experiences on buying from certain vendors. I expect that both of these are crucially important to the demand in this market. Because there are virtually no other places on the Internet for consumers to communicate with each other (largely due to criminality concerns), vendor ratings, past reviews, and the forums are the only ways for potential customers to collect information about a product or seller.

It is clear, especially since reviews are required of every past customer, that this will be one of the primary channels for customers to collect information on whether a vendor is selling high or low quality products, and whether or not they are committing fraud. Ex ante, though, it is not obvious that customers would highly value positive or negative reviews of vendors on forums. Since forum posts can be written by anyone, not just past customers, a consumer might view a vendor review on a forum as ``cheap talk'' since it is costless for a single user to send an arbitrary message on the forums, and thus they may ignore this information entirely. And so, especially since the author's utility is unlikely to be determined by the potential customer's purchase decision, the outcome may be a babbling equilibrium where forum messages are entirely noninformative \cite{farrell1996cheap}. At the same time, one might imagine there are sufficient incentives for forum members to accumulate social capital among their peers and obtain a reputation to be a member in good-standing with the community (See Wasko \& Faraj \cite{wasko2005should} for an investigation of incentives for knowledge sharing in an Internet forum setting). Given the limited means for consumers to collect information, I hypothesize that in this marketplace, sentiment of vendors displayed on forums will have an influential role in customer demand. In addition, I propose the following hypotheses for how consumer demand will depend on differing sources of vendor sentiment, which I test later in the paper:
\begin{itemize}
	\item \textbf{Hypothesis 1}: Customers will value vendor sentiment more highly as evidence accumulates; that is, if there are more reviews, all else equal, the customer, in a Bayesian fashion, will value the mean vendor sentiment more (relative to their prior) when formulating posterior beliefs.
	\item \textbf{Hypothesis 2}: Customers will value vendor sentiment more highly from experienced customers versus inexperienced customers. By experienced customers, I mean customers that in some sense have a longer history of interactions on the marketplace from which to draw from. Given this, experienced customers are more likely to be well informed, and so their input should be more heavily weighted by customers making purchase decisions.
	\item \textbf{Hypothesis 3}: Customers will value vendor sentiment from peer communication less if product is a ``no finalize early'' purchase. The reason for this hypothesis stems from the disclosure literature. Those listings with a no finalize early flag are credibly committing quality disclosure of their product within a neighborhood of its true quality, since customers can themselves assess the quality of the product before finalizing the purchase. When this quality disclosure is credible, a sophisticated buyer and seller will be able to communicate the full information quality in equilibrium \cite{milgrom1986relying}. As a result, vendor sentiment will not be as important for the demand of no finalize early products; consumers should be able to rely on the vendor's own provided information on the product listing page, compared to those products which may require customers to finalize early.
	\item \textbf{Hypothesis 4}: Customers will find the review sentiment for a product on its listing page to be more important than sentiment for the vendor's other products. I hypothesize that reviews on the product page are more tailored or relevant towards the purchase of this product. For example, reviews of other products from the same vendor may be informative on how a vendor's customer service is, but may not contain information about the quality of the specific product a consumer is interested in purchasing. Therefore, product demand will be more impacted by review sentiment for the specific product, rather than sentiment about the overall vendor (as captured by review sentiment on all other product listings by the vendor).
\end{itemize}



\section{Data}

For my analysis, I use three datasets concerning the anonymous market for drugs on the Internet. All of these datasets were generously uploaded for public usage by Gwern Branwen \cite{gwern}.

\begin{figure}
\centering
\caption{Volume of the major marketplaces on the Darknet}
\label{fig:marketsizes}
\includegraphics{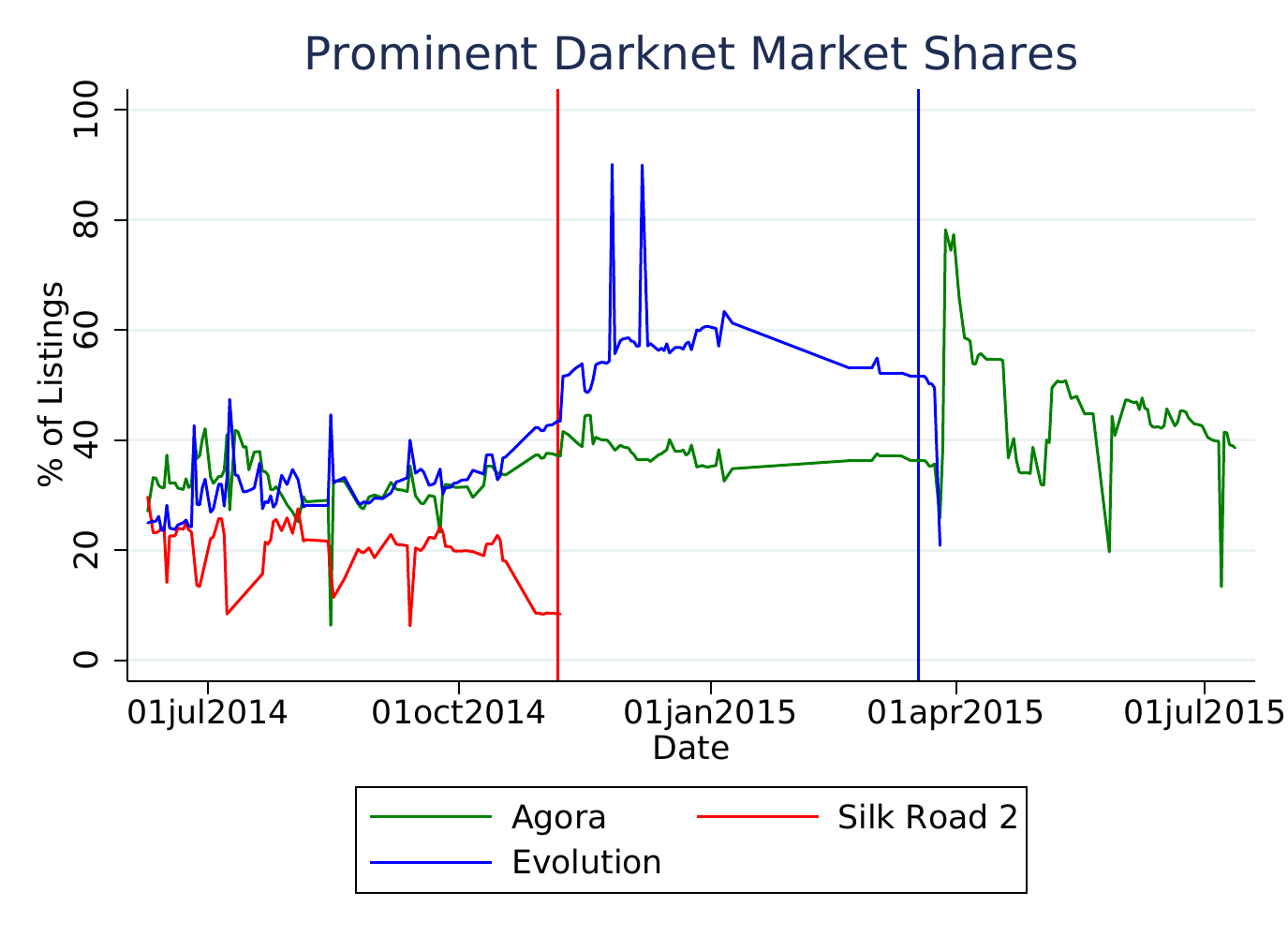}
\end{figure}

The first dataset is comprised of weekly, \textit{complete}, sets of listings available on all Darknet marketplaces from a central Darknet search engine named GRAMS. GRAMS interacts with the API of major marketplaces on the Darknet to obtain a complete set of listings on each marketplace. However, the information GRAMS provides is more basic than that contained in some of the datasets I describe below, a trade-off that must be weighed against its relative completeness. GRAMS has for each item the description, price, and seller name, along with a field for which Darknet market (Silk Road 2, Evolution, Agora, etc.) it is being sold on. Namely, it does not contain any sales data or review data for each listing. I use the GRAMS dataset as a sort of validation set to compare with my incomplete HTML marketplace data, as well as for testing vendor responses to changes in sentiment on an extensive margin.
\\ \indent I use some of the evidence in this GRAMS dataset to determine which marketplace websites would be best to study. Figure \ref{fig:marketsizes} shows the percent of listings (from the complete GRAMS data) on the Darknet from three of the largest marketplaces on the Darknet from 2014 to mid 2015.  The sample period can be broken up into three phases; before The Silk Road 2 went offline (red line), before Evolution went offline (blue line), and after. On November 6th, 2014, The Silk Road 2 server (started as a successor to the infamous original Darknet market, The Silk Road) was seized by U.S. customs authorities and brought offline \cite{sr2bust}. It did not enjoy as large of a market share as its predecessor, partly due to a loss of a first mover advantage, and partly due to its buggy interface. Evolution, on the other hand, enjoys the largest market share during the sample period I later focus on (the year 2014). Known for its high degree of security (unlike Silk Road 2), Evolution suddenly, without notice, went offline on March 14th, 2015, to the bewilderment of it's users. It was confirmed by moderators on the website that the owners of the site ``cashed out'' and stole the asset holdings of vendors currently in escrow  on the website, estimated to be worth \$12 million USD \cite{evoamount}.Because of the abrupt and peculiar circumstances surrounding the closing of the two other markets, I chose to not include these as the main object of study for this paper. Compare this to the shutdown of Agora, the other major market during this period, whose owners publicly announced in August 2015 that they were going to indefinitely take the website offline due to the increasing risk associated with running a Darknet website and security concerns. Everyone's asset holdings on the market were returned free of charge. Considering the sincere nature of it's exit from the marketplace , along with its relatively high market share, Agora seemed to be the best candidate market to study.

The second dataset I use consists of approximately biweekly  HTML scrapes of one of the Darknet's most prominent markets for drugs, Agora, from January 2014 to July 2015. The sample period I use for this research is data from the 2014 calendar year . These scrapes contain a rich amount of unstructured data on product listings on the market. Namely, they contain the complete product description provided by the vendor, an (optional) image of the product, the overall rating of the seller alongside the number of transactions the vendor has previously engaged in, shipping location, price, and partitions of the products into categories. Within categories, listings (theoretically) vary only on quantity and quality. In addition, each product page contains a history of reviews for the specific listing from previous customers, much in the style of online marketplaces such as Amazon and eBay. The web pages contain everything a consumer considering a purchase might see. Due to the instability of the TOR network that is required to connect to these online marketplaces, and semi-frequent DDOS attacks on the servers hosting the marketplace, many of these website scrapes are incomplete and often only cover a fraction of the listings on the marketplace on any given day. As a result, the observed time series of any individual product listing may be randomly right or left censored. For example, I might observe a listing in November of 2014  and never see it again, possibly because the vendor took the listing off the market soon after that scrape, or because the vendor left the listing on the site for an additional 2 months, but subsequent scrapes failed to arrive at the URL associated with that listing when crawling the website. The crawling procedure is described in \cite{gwern} and more or less follows a recursively defined random walk. Because of this, I can treat the listings I do observe as i.i.d. observations from the marketplace, since the random nature of the webcrawl ensures missingness will be unrelated to unobservable characteristics of the product.

The third and final dataset used in this paper is weekly scrapes of the Agora forums. On the Darknet, each marketplace typically has an accompanying forum where buyers and sellers can discuss offerings on the markets. Importantly, this is the main place where customers can discuss whether certain vendors are selling high or low quality items, or whether they are ``scams''. Due to the highly illegal and anonymous nature of the marketplace, it is very difficult to externally validate the quality of a listing. Importantly, these forums are virtually the only venue prospective buyers can go to for more information on the product. Consumers cannot go to typical online social networks for fear of being identified as a buyer or seller of illegal products. Since the Agora forums also require a Tor connection to anonymize users, it is a relatively safe area to discuss legitimate questions on products and sellers. The only other known place on the Internet where some discussions of Darknet markets occurs is Reddit, but the discussion is relatively limited for confidentiality concerns. The number of threads from 2014 in the Agora subreddit was 1,551; as a comparison, in this same period, the number of forumn threads in the Agora Forums was 52,058. Even among the relevant subforums I devote my analysis to, that are explicitly designated to contain topics directly related to the marketplace, have 10 times as many threads as this subreddit during this period.
Due to the concentrated nature of information on Darknet markets, the combination of the HTML of both the marketplace and the forum means that one can theoretically obtain a complete picture of the information customers would have had access to when deciding on a purchase. For this reason, I can completely characterize the information customers obtain from both formal (reviews) and informal (forum posts) means in this marketplace, which is what makes it such a uniquely interesting market to study.

The forum data has information on the subject text of individual thread, what topic the thread is in (General Discussion, Vendor Discussion,etc.) as well as the text of replies to the thread. In addition, each post is accompanied with information about the author, including the username of the author (which is the same as their username in the Agora marketplace), how active and experienced the user is, whether the author is classified as a seller by Agora, and overall favorability of the author's posts, as measured by ``karma'', much like favorability measures employed on social media websites such as Reddit.

Like the marketplace data, these HTML scrapes are often incomplete due to bandwidth limitations of the Tor network; however, because the forums are cumulative (posts are typically not removed after they are posted, and the forum website stores all threads in the history of its existence), I am more likely to observe the vast majority of posts in the data. It is for this reason that I limit my analysis of the Agora marketplace to the calendar year 2014, even though the data extends to July 2015. This will allow me to compile a more complete set of threads in my time period of interest, due to one of the most complete scrapes occurring in January 2015. Since threads are uniquely identified with an iterative integer (i.e. the first thread on the forums would have ID 1, the second thread has ID 2 etc.), I can also directly calculate the percentage of threads I am able to observe in the data. By this calculation, I observe 35,420 of 52,058 (68.04\%) of all the threads that have ever occurred on the forum. While some of the missing threads are due to the incompleteness of scrapes, it is likely that a substantial portion of missing threads are due to removal of a thread by moderators (for either being listed in the incorrect topic or violating the rules of the forum; spam is a common issue), so this should be considered a conservative lower bound on my coverage of the forum threads.


The basic unit of observation in this paper will be the time series of each listing on the marketplace. For each item, I observe every time it is sold, via the mandatory review, along with the date the transaction was completed (i.e. when the user reports they have received the product and completes the escrow transaction). From this, I can construct the number of transactions finalized each day, and use this as a proxy for the number of purchases each day. This follows the methodology used to measure sales in the modest literature that has studied Darknet markets. In order to reduce the impact of measurement error on the observed date of a sale, as well as to expedite computations, I aggregate the daily time series of sales data for each item to a weekly time series. I then incorporate vendor and item characteristics (price, average review rating, vendor rating, etc.) to construct a panel dataset of market listings over time, measured in weeks. As mentioned before, these time series will suffer from censoring because of the incomplete nature of the web crawls used to generate the dataset. I cannot be sure that an item that shows up for the first time in the web scrapes was not listed earlier on the marketplace, but simply missed by earlier scrapes. In this way, the listing time series' will suffer from random censoring, since the mechanism by which the series is censored (both left and right censorship are possible) is unrelated to any characteristics of the product itself. Because of this, the effect of the censoring on statistical inference will be limited, and I ignore it's impact on my estimates for coefficients.

Table \ref{tab:summstats} contains summary statistics of my panel dataset. As is evident in the table, product sales are relatively sparse (about 1 sale every 3 weeks per item). Across our sample period, we have 50,000 unique product from which to draw inference, sold by 2,000 different vendors. Both vendor ratings and item ratings are extremely high on average. Prices show extraordinary dispersion; this is mostly due to a couple of outlier listings that probably were mistakes by the vendor (possibly meant to put the price in USD terms); when the maximum price is excluded, the standard deviation is reduced by a factor of 10. To avoid these outliers altering the estimations, I log-transform prices whenever they are input into a regression equation. Table \ref{tab:summstats} also summarizes the \# of reviews by item / vendor, the average \# of mentions of a vendor on the forums, and the sentiment score variables we use later in the analysis. I now proceed to discuss how I take the rich text data from both reviews and forum threads, and turn it into useful covariates for regression analysis.
\section{Transforming Communication into Data}
\subsection{Vendor Sentiment in Text}
In the absence of verifiable information from sellers, buyers must seek alternative means of acquiring information on the quality or veracity of a particular product. I consider two avenues by which consumers acquire information. The first is through customer reviews. Since these reviews are required,
by design they should convey a relatively broad cross-section of past customer experiences. In practice, reviews are heavily biased upwards, resulting in a coarsening of information available to consumers. There is an immense pressure on Agora to give a vendor a 5-star rating for a product you purchase, in part because of the potential danger one risks by giving a distributor of illegal goods one's address (even if it is just a nearby P.O. Box). Many explicitly request it on their listing page, much like finalizing early. In fact, in my sample  97.7\% of reviews are 5 out 5 stars (one can rate a purchase from 0 to 5 stars).
Vendors care a lot about this rating since a vendor's overall rating (an unweighted average of reviews on their products)  is embedded into every active product listing by the vendor. In contrast, there are no widely distributed ``averages'' for the accompanying text required with each review. For this reason, I hypothesize that text, even in reviews with biased ratings, can provide valuable information to a prospective customer. In order to extract this information from the text, I use natural language processing methods to construct a ``sentiment'' variable associated with each review text that measures the relative positive or negative sentiment associated with a given product review, based on how certain words associate with customer sentiment of their purchase experience. After pre-processing the review text (removing stopwords, non alpha-numeric characters, as well as converting it to lowercase) using the Natural Language Processing Toolkit (NLTK) \cite{bird2009natural}, I decompose a review text into a matrix of word counts, and normalize the rows to sum to 1 (so the matrix consists of rows of word frequencies in an observed review). I then use  multinomial inverse regression (MNIR) \cite{taddy2013multinomial} to relate these frequencies to positive or negative product sentiment. Specifically, I assume that text for customer reviews is generated from a sequence of i.i.d. draws of words $w_i$ from a multinomial
\[w_i \sim Multinomial(\textbf{p}, m_i), p_j = \frac{e^{\alpha_j + \phi_j r_i + \epsilon_ij}}{\Sigma_k e^{\alpha_k + \phi_k r_i + \epsilon_ik}} \]
whose probability vector \textbf{p} is a  is determined by a latent measure of customer satisfaction of customer $i$, $r_i$. Here, $k$ indexes words across the entire vocabulary considered, while $\alpha_j$ and $\phi_j$ denote word-specific intercept and slope coefficients to relate each word's probability to $r$. I measure customer satisfaction with the star rating of a review each text is associated with. Even though this variable suffers from the bias and coarsening issues I just discussed above, this regression method should be able to identify strongly positive and negative text that accompanied all reviews, and give a better sense of whether, for example, the sample of 5-star reviews are really consistently associated with text as positive as the perfect rating given to vendors.

To calibrate this multinomial model's hyperparameters, I choose 10 regularization paths for the estimation (for computational tractability), and a hyperparameter choice of $\gamma=0$, which amounts to a complexity penalty equivalent to that found in lasso regression. I also choose to drop all words from the estimation procedure that do not appear at least 5 times in the set of $\approx 200,000$ reviews I study. I do this because it reduces the vocabulary I need to consider in the model(from about 23,000 to 6,000 unique stemmed words), which significantly eases the estimation computationally. At the same time, however, these ``scarce'' words are unlikely to be very informative about consumer sentiment and might result in extreme estimates for the score loadings $\boldsymbol{\phi}$ that are only based on a couple of observations. The choice of 5 words as the cutoff is an arbitrary choice I made that seemed reasonable in the tradeoff between limiting extreme knife-edge estimates for $\boldsymbol{\phi}$, and excessive censoring of text.

After estimating the multiomial model of word draws based on sentiment, I extract score loadings $\boldsymbol{\phi}$ associated with the star-rating for each word and apply these to the observed review text to construct the sufficient reduction projection scores $s_i= \Sigma_{k\in i} w_k\phi_k$, for each review. This score will capture all the available information in the text related to customer sentiment, in the sense that the review rating will now be orthogonal to the review text, conditional on the score variable \cite{taddy2013multinomial}.

Table \ref{tab:extremewords} shows the 20 most negative and positive words from the estimation in terms of their associated scores, alongside the observed frequency of each word and the average rating of reviews that contain each of these words. Both the negative and positive words are stems we might expect; 3 out of 20 of the most negative words allude to the product being a ``scam'', and the others mostly refer to either dishonesty on the part of the vendor or warn other customers to steer clear. In contrast, the positively scored words appear to signal praise for the vendor, or mention a positive experience with the delivery of the product. Notably, only 4 of the 40 words that occupy the extremes of the score spectrum actually explicitly refer to the quality characteristics of the product itself. The other high-scoring words appear to be concerned with the quality of the vendor themself. This may be because I am estimating sentiment from reviews across all types of product categories. Within each category, different words may be used to describe a product as low or high quality. I would not expect the word ``medibud'', for example, to be associated with a positive review of prescription pills. At the same time, all products listed on Agora suffer from common concerns and issues with delivery of the goods or reliability of a vendor. In this sense, my model is limited in that it will likely have difficulty identifying sentiment concerning ``product-specific quality'' as opposed to ``vendor-specific quality'' \footnote{one could imagine training a MNIR regression model separately or reviews in each sub-category, but this approach would be limited when comparing to the forum data}, but at the same time this identification strategy will prove useful when I analyze vendor quality sentiment in the forums. Another limitation of my approach will be that I do not account for the estimation error from the MNIR model in my later regression analysis when I use the scores $s_i$ as inputs; I take the output of the model as the true estimates. Correcting for this would involve boostrapping over the sample of reviews, re-estimating the MNIR model, and then performing all subsequent analysis using the random sample of reviews. Due to the intense computational requirements of both the MNIR estimation and the later regression analysis I perform, it is excessively cumbersome to do this analysis multiple times at this stage, and I ignore any error from the MNIR model. For this reason, we might expect standard errors to be biased downward in the later analysis.

\begin{figure}
\centering
\caption{Sentiment Scores by Review Rating}
\label{fig:sentimentbyStars}
\includegraphics[scale=.8]{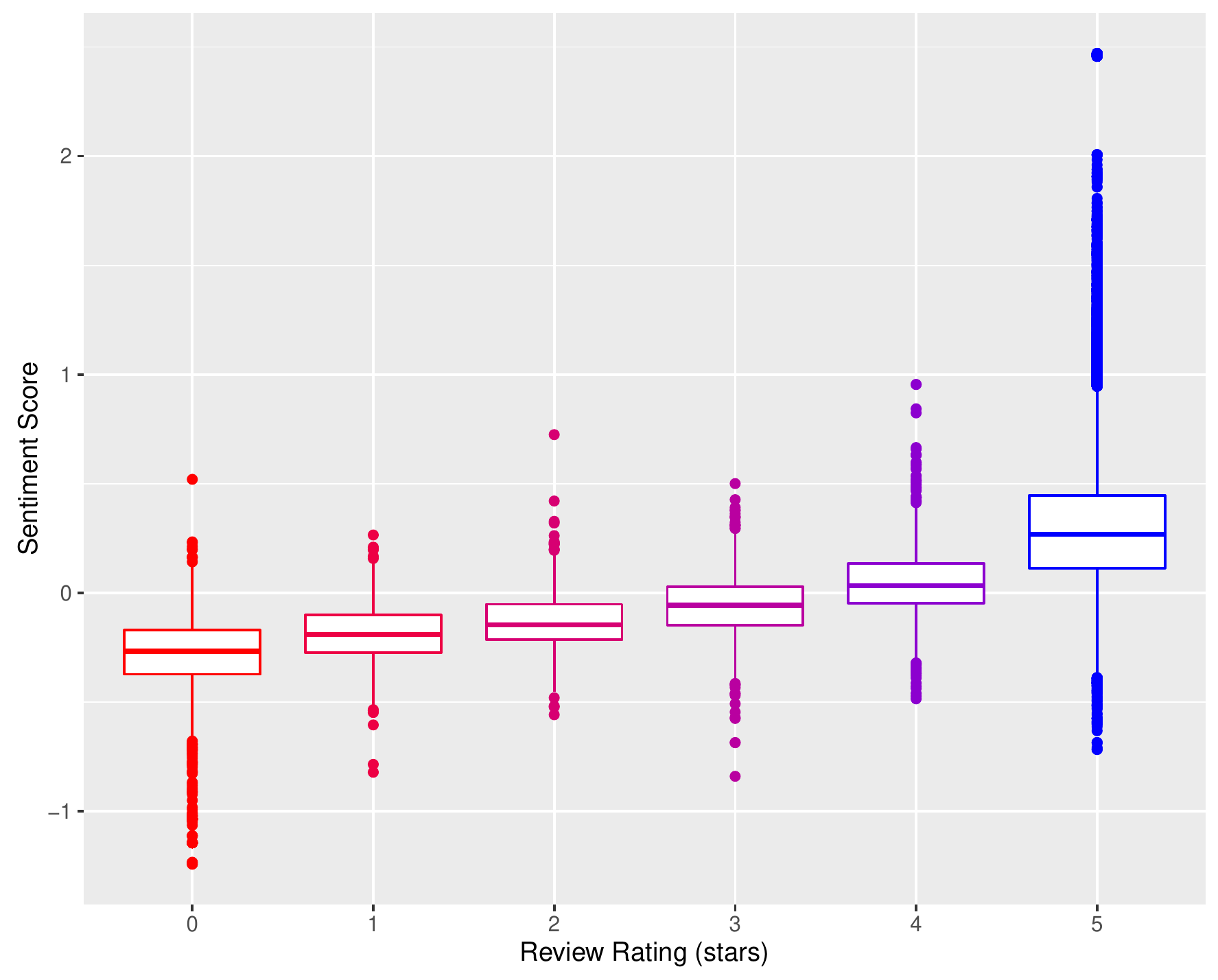}
\end{figure}

The sentiment score $s_i$ is the main covariate used in my analysis. Of the 6,013 score loadings I extract from the multinomial model, about 34\% of them have non-zero estimates for their values. This is in part due to my choice of a strict lasso penalty, rather than a semi-concave one, but in addition due to the limited variation in the review rating variable. The vast majority of reviews are 5/5 stars. Nonetheless, there is substantial variation in the score sentiment variables I construct from the text using the MNIR model. Only 7\% of reviews I consider have sentiment scores of zero, which would be the case when a review text is ``uninformative'' in regard to customer sentiment (as measured by not having any words with nonzero score magnitudes). Figure \ref{fig:sentimentbyStars} shows a boxplot of sentiment by the star rating of the review. While there is a clear monotonic relationship by number of stars given, there is substantial variation in sentiment within each star rating as well. In particular, we see that the most variation occurs in the 5-star category, as might be expected since it is by far the most populated. The range of the upper and lower hinges of the 5-star boxplot demonstrate that within 5-star reviews, there is enough variation in text sentiment to cover virtually the entire support of observed sentiment in 1-4 star reviews. This suggests that there is in fact a non-trivial amount of 5-star reviewers whose text aligns more with the language used in low rating reviews.

\subsection{Projecting the MNIR model onto Forum Data}
In addition to analyzing customer review text, I analyze the text of posts in each forum thread on the Agora forums. I now describe the process through which I analyze forum text as a complement to review text. The unit of measurement for analyzing forum text are forum posts which are classified as ``mentions'' that satisfy one of two requirements. The first is that the text in a post specifically includes a vendor's username in it's body. This is a direct mention. The second is if the post is located within a forum thread of which either:
\begin{enumerate}
	\item The title of the forum thread contains the vendor's username.
	\item The first post in the thread (the post by the originator of the thread) contains the vendor's username.
\end{enumerate}
The idea behind this latter identification strategy is that if a forum's title or first post has a vendor username, I assume the topic of the thread is the mentioned vendor, and specifically the vendor's quality. If a post mentions multiple vendors in the same body of text, for my main analysis, I drop the post, since it is impossible for me to evaluate within a forum post which positive or negative words are directed at which vendor. Later, as a robustness check, I rerun my baseline analysis by counting mentions of multiple vendors in a single post as separate but identical ``reviews'' of each individual vendor's quality. In practice this does not happen often. In order to reduce the noise I might introduce by including mentions that are not actually discussing with the quality of the vendor, I limit my analysis to forum posts within 4 sub-forums: General Discussion, Vendor Discussion, Product Offers, and Product Categories. The other sub-forums I exclude are: Referral Links, Newbie Section, Generic Randomness, German, Security discussion, Bugs, Philosophy, New features, and News. If forum posts are correctly categorized, it is unlikely that any of these excluded subforums would contain a vendor mention about the vendor's quality in relation to their products or customer service. Additionally, I exclude any vendor whose name is an English word, according to the Natural Language Toolkit's corpus of English words. I do this because it will be impossible to distinguish mentions that refer to the actual vendor in question or happen to just be using the English word the vendor is associated with. In my later analysis, to be consistent, I also throw out all reviews and product listings by vendors whose name is an English word as well, since these vendors may be discussed on the forums but my mention identification strategy would fail to identify them. About 5\% of the vendors in my data (the set of users who listed a product at least once in the listings dataset in 2014) have an English word as a name. I also remove one specific user, whose name is ``ebay'', who had posts associated with them in the forums that were actually discussing counterfeiting methods on the popular auction website eBay. Finally, I remove from my sample all posts by a vendor where they self-advertise (since consumers will presumably discount this information given the source). This constitutes a very small portion of all observed posts.

Using this definition of vendor mentions, I identify 61,679 unique posts that mention a vendor. Of these, 17,234, or  28\%, are a direct mention, and rest are posts in forum threads whose subject line includes a vendor. Some examples of the titles of these forum threads are ``californiadreamin vendor. legitimate or not? feedback?'', ``beware goingpostal, they are scammers!!!!'' and ``current whitelist of  opiate vendors (updated on the regular)''.

Like all choices, my choice of how to identify vendor mentions in the Agora forums is somewhat arbitrary and has its limitations. I cannot rule out that some threads might veer off-topic and discuss an entirely new subject, unrelated to the vendor's quality. Inversely, there may be some posts that are in fact reviewing a vendor's quality, but fail to name the vendor in their text body (perhaps a reply to a direct mention of a vendor). Additionally, there will be some posts I do not capture simply because a particular thread was never scraped during the data collection process.

\begin{figure}
\centering
\caption{Distribution of Informative Sentiment in Reviews and Forum Posts}
\label{fig:informedSentimentbySetting}
\includegraphics[scale=.8]{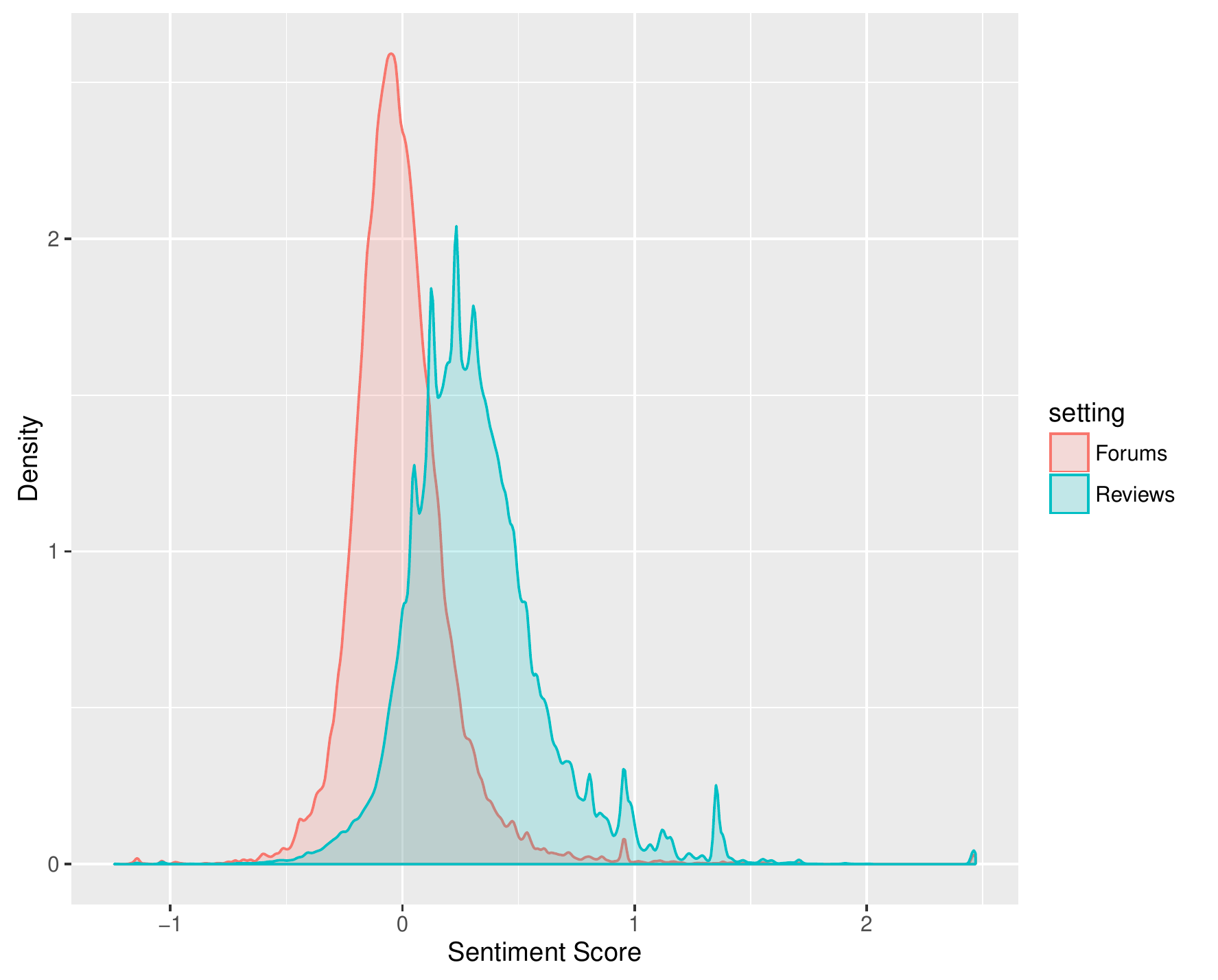}
\end{figure}

A primary interest of this paper is to compare the relative impact of formal and informal sentiment on vendor quality. I take all 61,679 posts flagged as mentioning a specific vendor on the Agora forums, and convert the text to a vendor sentiment variable, analogously to how I did in the review text setting. However, in this setting I have no labeled dependent variables; there is not an associated numerical rating of a vendor with each forum mention. So, I treat these forum posts as coming from the same population as the customer reviews, and apply the  MNIR model previously estimated on customer reviews to the forum posts. Specifically, I preprocess all post text as I did with reviews, treat each individual post as a unit of observation, and apply the score loadings $\boldsymbol{\phi}$ I estimated with the labeled review data to the sequence of words $w_p$ contained in each post, generating a vendor sentiment score for each forum post, $fs_p$. To my surprise, there is a smaller fraction of forum posts that are ``uninformative'' (i.e. $fs_p=0$); 3\% of the forum posts are uninformative, while about 7\% of formal reviews are uninformative. One can imagine that, even though posts mention a specific vendor, and are in one of the 4 appropriate subforums, some of these flagged posted are discussing an aspect of the vendor that is unrelated to consumer sentiment. There is a substantial amount of variation in the sentiment in forums as well. Figure \ref{fig:informedSentimentbySetting}  shows the density function of informative (non-zero) customer sentiment scores in both the set of forum posts and customer reviews. The density function that includes \textit{all} consumer sentiment is similar in appearance but with a large spike at zero, which makes the plot more difficult to interpret due to the scaling. The distributions of forum and review sentiment appear comparable in shape, but the distribution of forum sentiment is clearly shifted to the left of review sentiment. An interpretation of this is that sentiment on the forums is more negative towards vendors than on the formal reviews. This suggests that the forums may serve as an unrestricted setting where users can communicate their actual opinion on vendors more freely without the threat of blackmail that might accompany a very negative product review. Or it may be that the set of users who post on forums and those who actually purchase products simply differ in their attitudes or beliefs towards vendors.

Having generated scores $\{s_i\}$ and $\{s_p\}$ for reviews and forum posts, respectively, I (separately) standardize the non-zero scores  in each setting to have mean 0 and standard deviation 1\footnote{I specifically excluded the zero-score text from the mean and standard deviation calculations because standardizing these reviews would cause them to take on a non-zero value, which is difficult to interpret since I know these scores are zero because they have no informative text in their messages. note that overall forum and review sentiment variables will still have mean 0 after this transformation, but a lower standard deviation than 1}. Doing so eases the interpretation of coefficient estimates for later analysis, so that a 1 standard deviation increase in (informative) sentiment will result in an additive $\beta $ or multiplicative $e^\beta$ increase in the dependent variable I consider, depending on whether I use a linear or exponential mean model. For each vendor, I then construct an aggregate average vendor sentiment variable for the history of reviews of vendor $v$ up until week $t$:
\[\widetilde{s}_{v,t} = \frac{\Sigma_{i=1}^{n_v} s_{i,v} \mathbbm{1}_{\{t(i,v) < t\}}}{\Sigma_{i=1}^{n_v} \mathbbm{1}_{\{t(i,v) < t\}}} \]
 where $t(i,v)$ maps a review by consumer $i$ on a product sold by vendor $v$ to the week it was posted. Analogously, I construct for each vendor an average sentiment variable from forum mentions; that is, the mean of forum vendor sentiment scores $fs_{p,v}$ until week $t$:
 \[\widetilde{fs}_{v,t} = \frac{\Sigma_{p=1}^{m_v} fs_{p,v} \mathbbm{1}_{\{t(p,v) < t\}}}{\Sigma_{p=1}^{m_v} \mathbbm{1}_{\{t(p,v) < t\}}}\]
  where $m_v$ is the total number of forum mentions of a vendor in the sample period. Additionally, I construct total counts of the number of vendor's product reviews / vendor mentions up until time $t$ in the same manner (denoted $\widetilde{n}_{v,t}$ and $\widetilde{m}_{v,t}$, respectively). I choose to construct review sentiment variable's at the vendor level (rather than at the item level) so that average vendor sentiment between the two settings (reviews and forums) can be more directly comparable. Since discussion on forums mostly concerns vendors, rather than the specific items they sell, identifying specific mentions of products would have been challenging and probably yielded significantly less forum posts to work with. Because some vendors may not have a history of reviews or forum mentions at any given point in time, I always include a missing dummy when including these variables in a regression analysis.

Tables \ref{tab:reviewSentCorr} and \ref{tab:forumSentCorr} contain regression output from regressing (standardized) sentiment in each setting on virtually all observables available. We see, in the review setting, that after adding product reviewed and review week fixed effects, there is monotonically increasing relationship between sentiment scores and stars given in the review (as should be the case). In addition, other interesting relationships include that buyers with a longer transaction history are overall more positive in their review sentiment. Also, the buyer's personal rating (vendors may voluntarily review buyers who buy from them) does not display a clear relationship, except that those without a perfect 5-star rating give more negative text feedback. As may be expected, reviews on no finalize early products are more positive than those that ask customers to finalize early (this effect is estimated from listings that within their listing time switch from no finalize early to finalize early, or vice versa). In regards to forum sentiment, we see that experienced forum members are less likely to be positive in their sentiment. This descriptive result is suggestive of experienced vendors being sophisticated in how they share information with the community: they may know to be ``good'' to vendors in their product reviews, yet on forums, they are less worried about how a vendor might punish their negative report, and can speak more openly. We find no correlation, after controlling for a variety of fixed effects (username of the poster, vendor they are mentioning, and week the post is published), of karma, a social media similar in spirit to Facebook likes, relating to how positive or negative forum post sentiment is.

I also include in these tables one model regressing number of words  in a review or mention on observables. We see that non-5 star reviews on average  have ~3 more words, which may be expected since a review $<$ 5 stars represents a deviation from the default and so may indicate more thought is being given to these reviews. We also see some evidence of review fatigue: more experienced customers have, on average, less words in their reviews than newer buyers on the market.
\subsection{Item Listing Text}
A characteristic of a product listing that does not cleanly enter a regression equation is the listing text description of the product, along with the title of the listing. It is easy to see that changes in the description can in fact cause changes in sales: perhaps by editing the text to be less misleading about the
product's quality, or making the title more engaging to consumers scrolling through a list of products when searching the Darknet. Unlike the previous section, where I was able to compare reviews against each other because of associated ratings, there are no labels on the listing text, so I cannot say which listings texts represent effective or ineffective advertisement of the product. Thus, I need an alternative methodology to MNIR in order to be able to successfully control for changes in listing text over the lifespan of a given product.
As before, I preprocess the text of listings and then transform the title and description text (separately) into word frequency matrices for all of the sample listings. If all the information in the text relevant to consumers is in these word frequency matrices, then I could simply include the word matrices as additional controls in my regression to capture possible changes in sales due to changes in the phrasing of the listing. However, due to the massive size of these word frequency matrices, (14,000 unique word stems in titles, and 71,000 unique word stems in descriptions), this is computationally infeasible. To circumvent this, I decompose the word frequency matrices into principal components (PCs) that can explain a large portion of the variation in the listing text data. By using  the lower-dimensional principal components as controls, I can approximately control for the word frequencies.
Due the size and sparsity of these word frequency matrices, I use truncated singular value decomposition (TSVD) to retrieve the initial PCs of each listing text description. Furthermore, I do this principal component decomposition on the ``joint'' title-description matrix that is obtained by merging the title and description word frequency matrices along observations, so that all rows now sum to 2, and each contain the frequencies of words within titles and descriptions separately. The motivation for this is that, rather than extracting, for example, the first 100 PCs of text and description separately, then including the 200 PCs in my regression specification as controls, there are likely to be strong correlations between words present in the title and words present in the item description. So doing a PC decomposition on the joint matrix will require fewer PCs to explain the same amount of variation in the data. I chose to not just lump the title word counts into the counts of description words since, on average, description text is much longer than the title text, so the title text would contribute much less to the estimated principal components. the joint matrix approach I use more equitably weights the principle components explaining variation in the title and description listing text.
\section{Estimation}

This paper limits its inference to the effects of our covariates of interest on expected product demand, which the Poisson distribution is found to be asymptotically robust to. This is the case even if the true underlying demand function differs from Poisson, as long as the true mean function shares the exponential mean functional form \cite{cameron2013count}.
An alternative interpretation is that the true underlying model of product demand has covariates that are multiplicative factors of product demand (as opposed to additive, as is necessary in the linear regression model). This is a fairly general assumption in this setting, and if satisfied, we can consistently estimate the multiplicative parameters $e^\beta$ and their standard errors, robust to misspecification. Given our large sample size (about 600,000 observations of weekly sales data at the item level), I am not too concerned with our estimates deviating greatly from their asymptotic counterparts.

\begin{figure}
\centering
\caption{Distribution of Sales}
\label{fig:sales_counts}
\includegraphics[scale=.8]{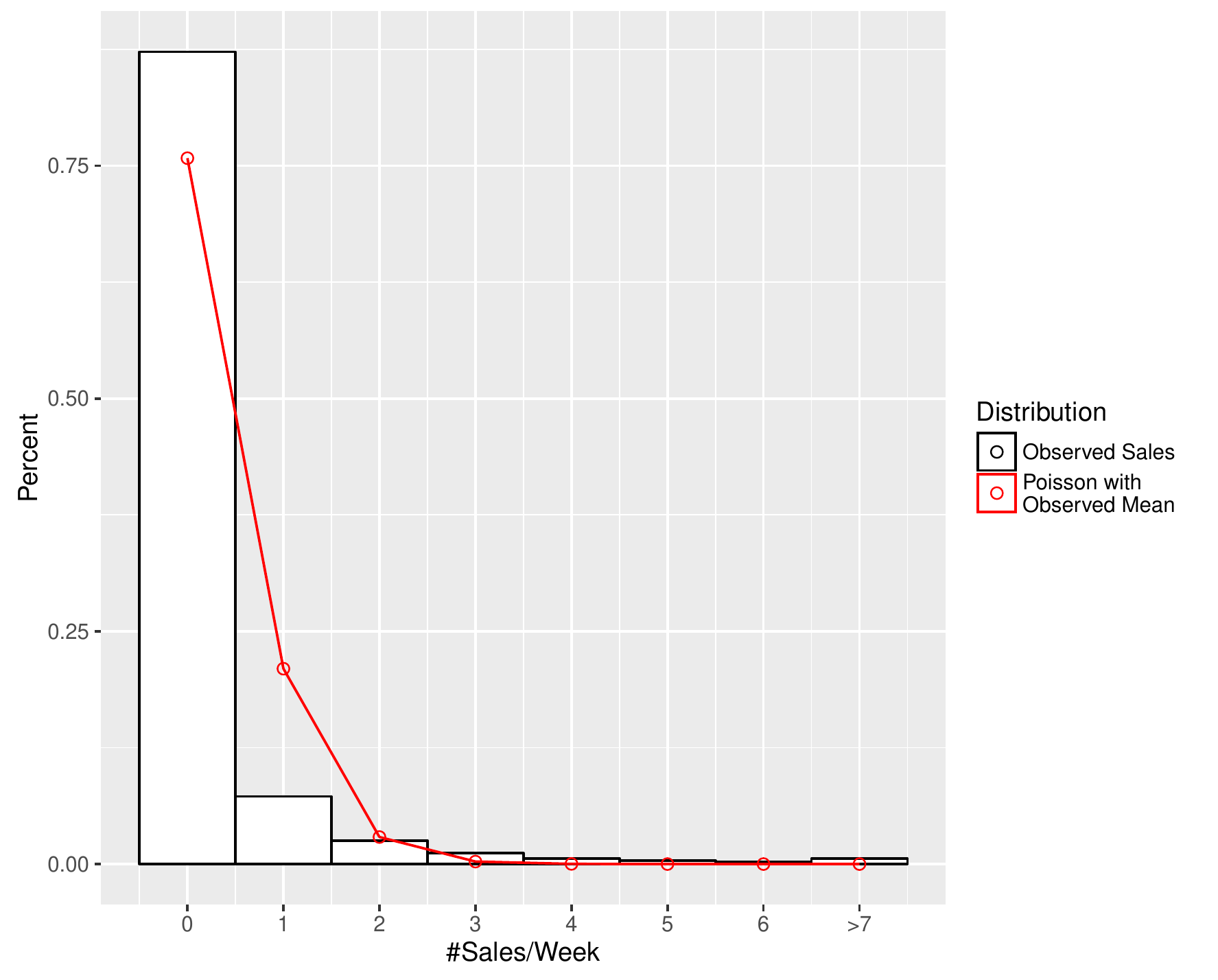}
\end{figure}

A simple histogram of sales per week by item listing (Figure \ref{fig:sales_counts}) clearly indicates that sales of individual products per week are exponentially distributed, as might be expected. In fact, when we compare the empirical distribution of observed sales against a Poisson distribution with mean parameter $\lambda$ equal to the sample mean of weekly item sales, we see that the Poisson distribution, even on the unconditional empirical distribution, appears to be a fairly good approximation. One notable feature of the data is that while it appears similar in shape to a Poisson distribution with the same mean, it also appears overdispersed, as is evident from the additional mass on the $0$ and $\geq7$ sales bins. Indeed, the sample variance is about 4 times as large as the sample mean (1.21 versus 0.28).

 Given information on sales, sentiment, and other product/vendor characteristics, I estimate the relationship between my covariates of interest and product demand via Poisson regression. Based on the unconditional distribution of sales, modeling weekly product sales as Poisson appears to be a reasonably good approximation. In addition to this, I prefer Poisson regression (as opposed to other count data regression models) due to its closed form solution to incorporating fixed effects, which allows me to non-parametrically control for unobservable time-invariant product characteristics.
My regression model for the baseline specification is the follows:
\begin{multline*}
\log(E[y_{j,t}|\mathbf{X}]) =  \beta_1 \widetilde{s}_{v,t} +\beta_2 \widetilde{fs}_{v,t}\\
+\alpha_1 \log(\widetilde{n}_{v,t}) + \alpha_2 \mathbbm{1}\{\widetilde{n}_{v,t}=0\}+
\alpha_3 \log(\widetilde{m}_{v,t}) + \alpha_4 \mathbbm{1}\{\widetilde{m}_{v,t}=0\} \\
+\mathbf{\gamma_1} \mathbf{X_{v,t}} + \mathbf{\gamma_2} \mathbf{X_{v,t}} +  + \mu_j + \delta_t
\end{multline*}
where $\mathbf{X_{v,t}}$ indicates a matrix of vendor controls, $\mathbf{X_{v,t}}$ indicates a matrix of product controls, $\delta_t$ and $ \mu_j$ are product and time fixed effects, respectively.
 I include a variety of controls to remove all supply-side variation available to consumers from the sales data when estimating my model of relating vendor sentiment to product demand. As I have stated before, by including variables for \textit{all} decisions made by the vendor that are observable to the consumer, I can ensure that a change by the vendor to, for example, flag the listing as no finalize early, is not confounded with the measured effect of vendor sentiment on demand. These include:
 \begin{itemize}
 \item Vendor Controls: the average vendor rating, a missing vendor rating indicator,and indicators for the past volume category the vendor selling the product is in. For example, in Figure \ref{fig:samplelisting}, the vendor's volume category is ``6$\sim$10 deals'' and their rating is 5/5.

 \item Item Controls: This includes an indicator for the listing being flagged as no finalize early, the natural log of the price of the item, the average item rating based on all reviews of that particular item, an indicator for zero item-specific reviews, and an indicator for the product category (absorbed by item fixed effects, but present when the regression is estimated with random effects). In addition, I control for the item description of the listing and the text title with the first 100 principal components of the joint word frequency matrix, which explain 1/3 of the variation in the item listing text data. As a robustness check, I re-ran my baseline specification with 300 PCs, explaining 1/2 of the variation, and found results to be nearly identical in my covariates of interest.\footnote{Results of these regressions are available by request.}
 \end{itemize}
 By also including time fixed effects (so indicators for each unique week in the dataset), I am able to implicitly control for any changes in aggregate market conditions, such as the stability of the TOR network, the Bitcoin exchange rate, and seasonal changes in sales (on both the demand and supply side). Through controlling all of these alternative channels that may effect sales, I can then cautiously interpret coefficients from regressing sales of a listing on vendor sentiment as responses in demand (since responses in supply that effect a consumer's choice are already conditioned on).

\section{Results}

\subsection{Baseline Specification}
I now introduce the main regression results of the paper. Table \ref{tab:baselineRegs} displays estimates from a Poisson panel regression of sales on vendor sentiment in both the forum and customer review setting, along with the (logged) \# of reviews \& mentions. The panel regression is done with respect to each product listing (so it includes product fixed effects), along with cluster-robust standard errors at the item level to allow for arbitrary correlations across observations of an individual listing. I iteratively add more controls column-by-column to show the persistence of the relationships that are estimated. The main specification of interest is column (5), where I control for all relevant supply responses, in addition to time fixed effects, so that the coefficients may be interpreted as demand responses to sentiment (i.e. elasticities).

The table shows multiplicative effects of a 1 unit increase in each regressor. Both measures of sentiment are averages of individual reviews/post sentiment that are normalized to have standard deviation 1 and mean 0 (among non-zero scores). The interpretation of the coefficient of, for example, Avg. Vendor Review Sentiment, is that an increase (decrease) in vendor sentiment among all reviews by 1 standard deviation leads to an 8.75\% increase (8.05\% decrease) in quantity demanded for the offered product. I find in the baseline regression that vendor sentiment on forums and reviews are approximately equal in their influence on demand- an average increase in 1 standard deviation of forum posts mentioning vendors leads to a 10.7\% increase in sales, which is not statistically significantly different from the impact of average vendor sentiment in reviews. Additionally, both have the expected sign - more positive vendor sentiment leads to higher demand. This verifies the hypothesis I had entering this paper- forum-based vendor sentiment has a substantial impact on consumer demand, and consumers in fact appear to value this information at least as highly as the information they glean from the text present in product reviews of a vendor. This is somewhat surprising, given the potential for mischaracterizations or lies in mentions on forums, and my somewhat crude method for identifying relevant forum posts. This underscores the potential importance of a designated space for Darknet market participants to freely exchange information on vendors.

Besides the sentiment measures, other variables for the most part have the expected sign: average item rating (on a 0-5 scale) is very positively associated with increased demand, price has a negative multiplicative (insignificant) effect on sales, and flagging one's post as no finalize early has a positive effect. Note that we cannot interpret the coefficient on price and no finalize early as effects on demand, since the response in sales to these changes could be endogenously driven by the vendor's decision to change the price / flag (i.e. low sales lead them to lower the price). Vendor rating, somewhat surprisingly, is insignificant when we include all of our additional controls. This may reflect the fact that the vendor review sentiment variable is a ``sufficient statistic'' for the vendor's average rating, and so its inclusion in the regression makes the vendor rating variable irrelevant. Conditional on scores $s_i$ , the review ratings $r_i$ should be orthogonal to the words contained in the review text. A more positive interpretation of this finding is that the star ratings in reviews across a vendor's products are informative insofar as they reflect the information contained within the text; no additional demand-relevant information is present in the numerical rating. This cannot be completely true however, given the strongly positive association between sales and the product-specific average rating. However. it may be the case that information on vendor-wide attributes (such as communication, customer service) is transmitted exclusively through review text, while product attributes (such as quality of the good) may not be completely conveyed in the review text. The negative relation of total count on past reviews with future sales is simply capturing the limited stock of vendors. Since past reviews exactly coincide with past sales, vendors with a fixed supply will mechanically have to sell less as past sales accumulate.

\subsection{Differential Effects}

Having established the importance of overall average vendor sentiment on product demand, I now look to uncover some heterogeneous effects of vendor sentiment on product demand. Specifically, I test the hypotheses I proposed in section \ref{consumermodel}. These hypotheses are formally tested in Table \ref{tab:diffEffectRegs}, where I control for all the supply variation I considered in column (5) in Table \ref{tab:baselineRegs}.
\begin{itemize}
	\item \textbf{Hypothesis 1}: This is tested in column (1) of Table \ref{tab:diffEffectRegs}. I test this by interacting average review/forum vendor sentiment with logged \# of reviews/posts. The coefficient on (uninteracted) average review/forum sentiment is now completely insignificant, while the interaction term is highly significant for both. This aligns nicely with a Bayesian framework of consumers updating their perceptions of vendors based on a stream of incoming reviews: consumers move their prior more (change their demand) when the observed sentiment in reviews/forums is associated with a higher sample size.
	\item \textbf{Hypothesis 2}: This is tested in column (2) of Table \ref{tab:diffEffectRegs}. I take all reviews, and split them into two groups: those by buyers who have bought at least 10 times on Agora, and those who have bought less (buyer past volume is observed on every review). This corresponds to an approximately median split of the sample of reviews in my dataset. Similarly, for forum posts, I split them into posts by forum members by ``ranking'', a title given to active/inactive users. My split is approximately based on whether or not a user has $\geq$ 200 posts\footnote{See \href{https://bitcointalk.org/index.php?topic=178608.msg2514705\#msg2514705}{here} for an official discussion of the rankings system by the same administrator of the Agora forums.}, where those with $\geq 200$ posts classified as experienced members (this is about 25\% of the sample, no cut closer to the median is available). Examining the regression results, I find that, to my surprise and contrary to my initial hypothesis, it appears that vendor sentiment in reviews from \textit{inexperienced} buyers is more influential on consumer demand (relative to experienced buyers). On the other hand, there is no observed heterogeneity based on my classification of experience in the forum setting. The ``P-value'' fields at the bottom of column (2) of the table shows the p-values from tests of equality of the coefficients on average sentiment between experienced and inexperienced users, done separately for both the review and forum setting. The non-differential effect of experience in forums appears to be a consequence of the way I measure experience (\# of past posts). This may not be particularly informative of a forum member's quality. The result that inexperienced members communicate vendor sentiment that is more influential on product demand is a curious one when taken at face value. However, there is a possible explanation. Directing our attention back to column (4) of table \ref{tab:reviewSentCorr}, we see some evidence that experienced and inexperienced buyers are giving different types of reviews, and this might be the driver of the result. Namely, less experienced buyers are, on average, inputting more words into each of their reviews, and their reviews are overall less positive than their more experienced counterparts. It may then be the case that inexperienced buyers are, perhaps naively, submitting more thoughtful reviews that are in fact more informative, while their experienced counterparts are foregoing the additional cost of giving an informative, possibly negative review that might agitate the vendor. It is worth reiterating that this argument is based on indirect evidence, and without explicitly controlling for each individual review's characteristics (which is difficult when we are aggregating reviews into an item panel framework), it cannot be verified.
	\item \textbf{Hypothesis 3}: This is tested in column (3) of Table \ref{tab:diffEffectRegs}. As before, I test whether the coefficient of average sentiment on finalize early versus no finalize early listings are equal, and display the p-value in the cells below. In both the forum and review settings, I find evidence that those listings that allow finalize early are impacted about twice as much by vendor sentiment communicated in reviews and forum posts. Although the coefficient on the interaction of finalize early and review sentiment is not significantly different from its no finalize early counterpart (p-value = 0.2), it is significantly different from zero. This confirms the hypothesis that those potential customers of products with a commitment to not finalize early will be able to take what the vendor reports in the listing description more at ``face-value''. It is particularly noteworthy that, despite the uninteracted no finalize early indicator entering positively and significantly in the previous table, its interaction yields results pointing to less importance when analyzing the vendor sentiment available to customers (i.e. I am not capturing the fact that those vendors with no finalize early listings are more positively reviewed than those with finalize early listings).
	\item \textbf{Hypothesis 4}: This is tested in column (4) of Table \ref{tab:diffEffectRegs}. For each listing, I isolate the set of reviews for that product, and compare the impact of the average review sentiment these reviews have to the average sentiment for all other reviews given of a vendor up until time $t$. To my surprise, I find that the average of ``other'' product reviews is 5 times as influential on customer demand (2\% increase versus 10\% increase)! However, this has an explanation besides product demand being more influenced by reviews for other products. Since a vendor is likely to list multiple products, there are, on average, many more reviews for a vendor's other products than any individual listing. As a result, when one controls for the additional information from sample size conveyed in average other product sentiment by interacting with logged \# of reviews (analogous to the procedure done in column (1), but this time separately for product reviews and other product reviews), the two produce similar coefficients that are not statistically significantly different from each other\footnote{Regression results of this interaction are available upon request}. This column, then, serves to reaffirm Hypothesis 1, that consumers value the signal sent by average customer sentiment more when it is accompanied by a larger sample of reviews, when formulating their posterior on vendor quality.
\end{itemize}
Overall, the largest takeaway from these differential effects is that consumer demand is significantly more influenced by vendor sentiment on the web when one increases the sample from which this average comes from. In this sense, there are definitely ``increasing scales'' to positive vendor sentiment if it continues to be perpetrated by more and more customers.
\subsection{Vendor Response on the Extensive Margin}
While I have shown that within the lifespan of a product listing, vendor changes in product characteristics can be appropriately controlled for, there may be other forms of response by the vendor. Namely, it may be the case that a vendor responds by pulling a listing altogether from the marketplace due to poor sentiment , and this may be distorting my interpretations of coefficients on vendor sentiment as demand responses. To investigate this potential confounder, I regress the total number of listings produced by each vendor on the sentiment variables discussed before, while controlling for vendor attributes. If it is found that vendors are not responsive to customer sentiment in text on the extensive margin, then it is unlikely this will be a significant confounder. Because I am now interested in evaluating weekly number of total listings, I can draw upon the GRAMS dataset in addition to the Agora scrapes. The GRAMS data will contain the uncensored count of listings for each vendor on Agora, so it should produce the more precise estimate of the true relation between vendor sentiment and number of posted listings on the Darknet.

Table \ref{tab:venlistings} shows the regression output of a Poisson regression of listings count on sentiment variables. I include a version of both fixed and random gamma vendor effects, and I draw my dependent variable from both the GRAMS and Agora HTML datasets. As is evident, there is no statistically significant vendor response in the number of listings by a vendor to text sentiment. Across the board, the regression using GRAMS data show zero effect of sentiment in both forums and reviews on a vendor's number of listings. And while the coefficient on vendor review sentiment on the Agora scrapes is significant at the 10\% level, it in fact moves in the \textit{opposite} direction one expects. The interpretation of the coefficient is that if positive sentiment about a vendor in reviews increases, they actually offer less products, rather than more. Given this counterintuitive coefficient, along with the estimated effect of zero everywhere else, and the noisy nature of the Agora data (since product counts may be incorrectly underestimated due to the web crawl timing out), I discount this finding as an outlier not reflective of the true relationship.

\subsection{Robustness Checks}
Here, I discuss some validity checks performed to confirm my results are not knife's edge in nature and survive a broad set of specifications.
To begin, Tables \ref{tab:baselineRegs_re} \& \ref{tab:diffEffectRegs_re} are the same specification as \ref{tab:baselineRegs} \& \ref{tab:diffEffectRegs}, discussed above, but this time estimated with random gamma-distributed effects instead of non-parametric fixed effects. We see that in this specification, qualitatively results are identical, and the multiplicative effects are on average slightly greater in magnitude. Nothing else substantially changes. In addition, we see that the $\alpha$ parameter of the
regression, the multiplier for the variance of the conditional Poisson distribution, is significant and positive. In fact, this specification (Poisson with gamma random effects) is equivalent to a negative binomial regression, which allows for overdispersion in the count data, and so leads to more efficient estimation when overdispersion is present \cite{cameron2013count}. These tables demonstrate that the main results of the paper are not a function of the Poisson regression assumption that the mean and variance of the conditional distribution are equal. \\ \indent
I also include a version of the regressions present in Tables \ref{tab:baselineRegs} \& \ref{tab:diffEffectRegs} estimated via linear regression (maintaining product fixed effects), in Tables \ref{tab:baselineRegs_linear} \& \ref{tab:diffEffectRegs_linear}. While a linearly additive model for non-negative count data is questionable, these regressions also yield coefficients that are qualitatively similar to those estimated in the tables discussed above. They provide evidence that the estimates I find are not a result of the maximum-likelihood estimation procedure getting stuck in a local mode due to the high dimensionality of the coefficient parameter $\beta$ (I have about 180 different regressors, excluding fixed effects, in my main specification) since OLS yields a closed form solution for the global optimum.

Finally, in Table \ref{tab:baselineRegs_allposts}, I estimate my main specification using the non-exclusive measure of forum mentions. Instead of limiting mentions to those that include only one specific vendor, I assign the posts that mention multiple vendors to each individual vendor as duplicates. The estimated effect is lower under this classification, possibly due to increased measurement error, but nonetheless significant at the 5\% level.

\section{Conclusion}
I present evidence in this paper that communication between marketplace participants - as captured by sentiment variables on vendor quality- is an important influence of demand for products. While that result is not particularly groundbreaking, I find that consumer demand is equally influenced by communication on both formal and informal networks - namely, product reviews versus community forums. This may come as a surprise since well-regulated product reviews should be more reliable, but the evidence presented here suggests that discussion even in unstructured settings like forums should be an important determinant of product demand. In addition, I find some empirical evidence that a vendor's ability to commit to disclosure, by flagging their listings as no finalize early, dampens the effect of communication on demand. Furthermore, I find strong evidence that product demand is more responsive to customer communication as the number of messages grows, as may be expected in a Bayesian updating framework.

There are some limitations to interpreting these results, however. Since my dataset is based on imperfect web crawls of the Darknet, I cannot perfectly observe when a change in product characteristics occurs. While this missingness in random, it may be the case that I am not perfectly controlling supply side variation, and so cannot completely attribute the effect of sentiment on sales to demand. By aggregating to the weekly level, I attempt to minimize the influence of this, but also introduce new errors: namely, that some customers who bought during a week may have been exposed to a different listing page, depending on when they purchased and when a vendor updated their product page.
This paper looks only at the effects on the Agora marketplace, but Agora is functioning alongside other Darknet markets, and their may be important cross-marketplace effects this paper does not explore. Additionally, I excluded from my analysis controlling for variation in the image shown on the product listing page. State-of-the-art machine learning algorithms exist to control for the information contained in an image, but these ultimately proved too cumbersome at this stage to include in this project.

Ultimately, this paper serves as a useful starting point for the study of a market that is exceptional for testing economic theory, due to its highly anonymous nature and independence from real-world regulations, but more work on the economic properties of Darknet markets should follow.

\bibliographystyle{plain}
\bibliography{citations}

\clearpage
\pagebreak

\clearpage
\pagebreak

\begin{table}
\centering
\caption{Summary Statistics of Weekly Item Panel Dataset}
\label{tab:summstats}
\input{summstats}
\begin{flushleft}
Means of variables reported in cells. Standard deviations in parentheses for continuous variables.
Sample means and standard deviations calculated at the Item-by-Week observation level, as in the constructed panel dataset used for the final analysis.
\end{flushleft}
\end{table}

\input{extremewordstable}

\begin{table}
\centering
\caption{Correlates of Review Sentiment}
\label{tab:reviewSentCorr}
\input{reviewCorr_unigram}
\begin{flushleft}
OLS estimates reported from regressing sentiment and number of words of a review on observables. Informative sentiment is standardized to have mean 0 and standard deviation 1.
Robust standard errors reported in parantheses in column (1), and listing clustered standard errors reported in parantheses in columns (2) to (4). *,**,*** denote significance at the 10, 5, and 1 \% levels, respectively.
\end{flushleft}

\end{table}

\begin{table}
\centering
\caption{Correlates of Forum Sentiment}
\label{tab:forumSentCorr}
\input{forumSentCorr_unigram}
\begin{flushleft}
OLS estimates reported from regressing sentiment and number of words of a post on observables. Informative sentiment is standardized to have mean 0 and standard deviation 1.
Robust standard errors reported in parantheses in column (1), and author clustered standard errors reported in parantheses in columns (2) to (3). *,**,*** denote significance at the 10, 5, and 1 \% levels, respectively.
\end{flushleft}

\end{table}

\begin{sidewaystable}
\centering
\caption{Impact of Consumer Text Sentiment on Product Demand}
\label{tab:baselineRegs}
\input{baselineDemandRegs_fe}
\begin{flushleft}
Estimates reported from a Poisson regression with product listing fixed effects and cluster-robust standard errors. Exponentiated coefficients $e^\beta$ reported. Exponentiated standard errors calculated using the delta method. Vendor controls include  vendor rating, a missing vendor rating indicator, and indicators for total volume of trade vendors have previously engaged in. Item Controls include the first 100 principal components of item listing text, a no finalize early indicator, logged price, average product rating, and a missing product rating indicator. Time FE are unique indicators for every week in the sample. These regressions also include indicators for zero reviews/mentions (not shown). *,**,*** denote significance at the 10, 5, and 1 \% levels, respectively.
\end{flushleft}
\end{sidewaystable}

\begin{sidewaystable}
\centering
\caption{Differential Effects of Consumer Text Sentiment on Product Demand}
\label{tab:diffEffectRegs}
\input{diffEffectsDemandRegs_fe}
\begin{flushleft}
Estimates reported from a Poisson regression with product listing fixed effects and cluster-robust standard errors. Exponentiated coefficients $e^\beta$ reported. Exponentiated standard errors calculated using the delta method. Vendor controls include  vendor rating, a missing vendor rating indicator, and indicators for total volume of trade vendors have previously engaged in. Item Controls include the first 100 principal components of item listing text, a no finalize early indicator, logged price, average product rating, and a missing product rating indicator. Time FE are unique indicators for every week in the sample. These regressions also include log(number of vendor reviews), log(number of forum mentions), as well as indicators for zero reviews/mentions (not shown). *,**,*** denote significance at the 10, 5, and 1 \% levels, respectively.
\end{flushleft}
\end{sidewaystable}

\begin{table}
\centering
\caption{Impact of Sentiment On Vendor Listings}
\label{tab:venlistings}
\input{VendorListingRegs}
\begin{flushleft}
Estimates reported from a Poisson regression with product listing fixed effects and cluster-robust standard errors. Exponentiated coefficients $e^\beta$ reported. Exponentiated standard errors calculated using the delta method.Vendor controls include  vendor rating, a missing vendor rating indicator, and indicators for total volume of trade vendors have previously engaged in. Time FE are unique indicators for every week in the sample. *,**,*** denote significance at the 10, 5, and 1 \% levels, respectively.
\end{flushleft}

\end{table}

\appendix
\renewcommand{\thetable}{\thesection A\arabic{table}}
\setcounter{table}{0}

\begin{sidewaystable}
\centering
\caption{Impact of Consumer Text Sentiment on Product Demand (Random Effects)}
\label{tab:baselineRegs_re}
\input{baselineDemandRegs_re}
\begin{flushleft}
Estimates reported from a Poisson regression with gamma distribution random effects and cluster-robust standard errors. Exponentiated coefficients $e^\beta$ reported. Exponentiated standard errors calculated using the delta method. Vendor controls include  vendor rating, a missing vendor rating indicator, and indicators for total volume of trade vendors have previously engaged in. Item Controls include the first 100 principal components of item listing text, category of product indicators, a no finalize early indicator, logged price, average product rating, and a missing product rating indicator. Time FE are unique indicators for every week in the sample. These regressions also include indicators for zero reviews and zero mentions (not shown). *,**,*** denote significance at the 10, 5, and 1 \% levels, respectively.
\end{flushleft}
\end{sidewaystable}

\begin{sidewaystable}
\centering
\caption{Differential Effects of Consumer Text Sentiment on Product Demand (Random Effects)}
\label{tab:diffEffectRegs_re}
\input{diffEffectsDemandRegs_re}
\begin{flushleft}
Estimates reported from a Poisson regression with gamma distribution random effects and cluster-robust standard errors. Exponentiated coefficients $e^\beta$ reported. Exponentiated standard errors calculated using the delta method. Vendor controls include  vendor rating, a missing vendor rating indicator, and indicators for total volume of trade vendors have previously engaged in. Item Controls include the first 100 principal components of item listing text, category of product indicators, a no finalize early indicator, logged price, average product rating, and a missing product rating indicator. Time FE are unique indicators for every week in the sample. These regressions also include log(number of vendor reviews), log(number of forum mentions), as well as indicators for zero reviews and zero mentions (not shown). *,**,*** denote significance at the 10, 5, and 1 \% levels, respectively.
\end{flushleft}
\end{sidewaystable}

\begin{sidewaystable}
\centering
\caption{Impact of Consumer Text Sentiment on Product Demand \\ (Including Multi-Mention Forum Posts)}
\label{tab:baselineRegs_allposts}
\input{baselineDemandRegs_fe_allposts}
\begin{flushleft}
Estimates reported from a Poisson regression with product listing fixed effects and cluster-robust standard errors. Exponentiated coefficients $e^\beta$ reported. Exponentiated standard errors calculated using the delta method. Vendor controls include  vendor rating, a missing vendor rating indicator, and indicators for total volume of trade vendors have previously engaged in. Item Controls include the first 100 principal components of item listing text, a no finalize early indicator, logged price, average product rating, and a missing product rating indicator. Time FE are unique indicators for every week in the sample. These regressions also include indicators for zero reviews/mentions (not shown). *,**,*** denote significance at the 10, 5, and 1 \% levels, respectively.
\end{flushleft}
\end{sidewaystable}

\begin{sidewaystable}
\centering
\caption{Impact of Consumer Text Sentiment on Product Demand (Linear Regression)}
\label{tab:baselineRegs_linear}
\input{baselineDemandRegs_fe_linear}
\begin{flushleft}
Estimates reported from a linear regression model with fixed effects and cluster-robust standard errors. Vendor controls include  vendor rating, a missing vendor rating indicator, and indicators for total volume of trade vendors have previously engaged in. Item Controls include the first 100 principal components of item listing text, a no finalize early indicator, logged price, average product rating, and a missing product rating indicator. Time FE are unique indicators for every week in the sample. These regressions also include indicators for zero reviews and zero mentions (not shown). *,**,*** denote significance at the 10, 5, and 1 \% levels, respectively.
\end{flushleft}
\end{sidewaystable}

\begin{sidewaystable}
\centering
\caption{Differential Effects of Consumer Text Sentiment on Product Demand (Linear Regression)}
\label{tab:diffEffectRegs_linear}
\input{diffEffectsDemandRegs_fe_linear}
\begin{flushleft}
Estimates reported from a linear regression model with fixed effects and cluster-robust standard errors. Vendor controls include  vendor rating, a missing vendor rating indicator, and indicators for total volume of trade vendors have previously engaged in. Item Controls include the first 100 principal components of item listing text, a no finalize early indicator, logged price, average product rating, and a missing product rating indicator. Time FE are unique indicators for every week in the sample. These regressions also include log(number of vendor reviews), log(number of forum mentions), as well as indicators for zero reviews and zero mentions (not shown). *,**,*** denote significance at the 10, 5, and 1 \% levels, respectively.
\end{flushleft}
\end{sidewaystable}

\end{document}

%% file: summstats.tex
\begin{tabular}{lc}
\hline\hline 
Variable & (1) \\ \hline
Number of Sales/Week per Item& 0.28 \\
 & (1.11) \\
Item Rating (0-5 Stars)& 4.86 \\
 & (0.57) \\
Price (BTC)& 3.64 \\
 & (925.79) \\
\# of Customer Reviews per Item& 3.50 \\
 & (11.09) \\
\# of Customer Reviews per Vendor& 123.74 \\
 & (247.64) \\
Vendor Rating (0-5 Stars)& 4.87 \\
 & (0.43) \\
Average Stdized Vendor Sentiment From Customer Reviews& -0.09 \\
 & (0.39) \\
\# of Mentions per Vendor on Forums& 37.43 \\
 & (199.63) \\
Average Stdized Vendor Sentiment From Forum Posts& 0.03 \\
 & (0.52) \\
\% of Items with Finalize Early& 56.68\% \\
\% Missing Item Rating (Zero Item Reviews)& 59.87\% \\
\% Missing Vendor Rating (Zero Vendor Reviews)& 5.75\% \\
\% Missing Vendor Forum Sentiment (Zero Mentions)& 27.17\% \\
\hline
Number of Items in Sample& 50012 \\
Number of Vendors in Sample& 2084 \\
Number of Weeks in Sample& 54 \\
Number of Unique Agora Market Scrapes in Sample& 109 \\
Number of Observations & 651143 \\
\hline \hline \end{tabular}

%% file: extremewordstable.tex
\begin{sidewaystable}[ht]
\centering
\begin{tabular}{|c|lcc|lcc|}
  \hline
Rank & Most Positive Words & Phrase Count & Avg. Review & Most Negative Words & Phrase Count & Avg. Review \\ 
  \hline
 1 & goto & 333 & 5.00 & thief & 13 & 0.00 \\ 
   2 & legend & 378 & 5.00 & enforc & 5 & 0.00 \\ 
   3 & ultra & 225 & 5.00 & ripoff & 11 & 0.18 \\ 
   4 & inde & 181 & 4.73 & scamm & 7 & 0.91 \\ 
   5 & western & 135 & 5.00 & dishonest & 5 & 0.80 \\ 
   6 & boss & 340 & 5.00 & scammer & 417 & 0.91 \\ 
   7 & medibud & 255 & 5.00 & unansw & 6 & 0.83 \\ 
   8 & dm & 224 & 4.86 & worthless & 7 & 0.86 \\ 
   9 & pl & 50 & 4.76 & avoid & 84 & 1.02 \\ 
  10 & thrown & 70 & 5.00 & scumbag & 5 & 1.00 \\ 
  11 & bone & 33 & 5.00 & bath & 5 & 1.67 \\ 
  12 & intens & 61 & 5.00 & steer & 5 & 1.00 \\ 
  13 & approv & 46 & 5.00 & bewar & 96 & 1.19 \\ 
  14 & meet & 37 & 5.00 & invalid & 6 & 1.17 \\ 
  15 & dab & 82 & 4.95 & liar & 11 & 3.46 \\ 
  16 & dnms & 56 & 5.00 & scam & 580 & 1.21 \\ 
  17 & homi & 49 & 5.00 & disabl & 6 & 1.33 \\ 
  18 & instrument & 70 & 5.00 & virus & 9 & 1.55 \\ 
  19 & ah & 45 & 4.92 & benzocain & 5 & 1.40 \\ 
  20 & deffo & 46 & 5.00 & stole & 14 & 2.45 \\ 
   \hline
\end{tabular}
\caption{Most Negative and Positive Words by Score from MNIR Estimation} 
\label{tab:extremewords}
\end{sidewaystable}

%% file: reviewCorr_unigram.tex
{\def\sym#1{\ifmmode^{#1}\else\(^{#1}\)\fi}\begin{tabular}{@{\extracolsep{4pt}} l*{4}{c}} \hline\hline
Dependent Variable: & \multicolumn{3}{c}{Standardized Sentiment} & Number of Words \\ \cline{2-4} \cline{5-5}
                    &\multicolumn{1}{c}{(1)}   &\multicolumn{1}{c}{(2)}   &\multicolumn{1}{c}{(3)}   &\multicolumn{1}{c}{(4)}   \\
\hline
Number of Words     &      -0.058***&      -0.059***&      -0.059***&               \\
                    &   (0.00045)   &   (0.00061)   &   (0.00060)   &               \\
0 Star Item Rating  &       -1.89***&       -1.66***&       -1.65***&        1.99***\\
                    &     (0.016)   &     (0.020)   &     (0.020)   &      (0.11)   \\
1 Star Item Rating  &       -1.52***&       -1.32***&       -1.32***&        2.74***\\
                    &     (0.025)   &     (0.031)   &     (0.031)   &      (0.21)   \\
2 Star Item Rating  &       -1.32***&       -1.16***&       -1.16***&        2.94***\\
                    &     (0.025)   &     (0.029)   &     (0.029)   &      (0.22)   \\
3 Star Item Rating  &       -1.09***&       -0.94***&       -0.94***&        2.47***\\
                    &     (0.017)   &     (0.021)   &     (0.021)   &      (0.14)   \\
4 Star Item Rating  &       -0.72***&       -0.61***&       -0.61***&        2.64***\\
                    &     (0.012)   &     (0.014)   &     (0.014)   &      (0.11)   \\
Buyer Has 1-5 Previous Deals&      0.0072   &       0.024*  &     -0.0024   &       -0.27***\\
                    &     (0.011)   &     (0.014)   &     (0.015)   &     (0.075)   \\
Buyer Has 6-10 Previous Deals&       0.066***&       0.078***&       0.047***&       -0.44***\\
                    &     (0.012)   &     (0.015)   &     (0.016)   &     (0.082)   \\
Buyer Has 11-25 Previous Deals&       0.090***&       0.099***&       0.066***&       -0.52***\\
                    &     (0.012)   &     (0.015)   &     (0.017)   &     (0.086)   \\
Buyer Has $ > $25 Previous Deals&        0.15***&        0.14***&        0.11***&       -0.55***\\
                    &     (0.013)   &     (0.016)   &     (0.017)   &     (0.090)   \\
Missing Buyer Rating&       -0.11***&      -0.074***&      -0.079***&       -0.33***\\
                    &    (0.0062)   &    (0.0072)   &    (0.0074)   &     (0.038)   \\
Buyer Rating $ < $ 3 Stars&      -0.089***&      -0.070***&      -0.079***&       -0.27***\\
                    &     (0.012)   &     (0.013)   &     (0.014)   &     (0.071)   \\
Buyer Rating $ \in [3,4) $&       -0.14***&      -0.094***&      -0.096***&      -0.079   \\
                    &     (0.027)   &     (0.031)   &     (0.031)   &      (0.15)   \\
Buyer Rating $ \in [4,5) $&      -0.066***&      -0.047***&      -0.047***&       -0.11*  \\
                    &     (0.011)   &     (0.012)   &     (0.012)   &     (0.057)   \\
No Finalize Early   &       0.080***&       0.076***&       0.074***&       0.020   \\
                    &    (0.0046)   &     (0.020)   &     (0.020)   &     (0.094)   \\
Constant            &        0.33***&        0.31***&       -0.21   &        5.78***\\
                    &     (0.013)   &     (0.021)   &      (0.19)   &      (0.89)   \\
\hline
Observations        &      197040   &      197040   &      197040   &      197040   \\
R-Squared           &        0.17   &        0.14   &        0.14   &        0.01   \\
Fixed Effects       &        None   &     Item FE   &Item+Time FE   &Item+Time FE   \\
Mean of Dependent Variable&       -0.00   &       -0.00   &       -0.00   &        6.58   \\
SD of Dependent Variable&        0.96   &        0.96   &        0.96   &        4.39   \\
\hline\hline \end{tabular}}

%% file: forumSentCorr_unigram.tex
{\def\sym#1{\ifmmode^{#1}\else\(^{#1}\)\fi}\begin{tabular}{@{\extracolsep{4pt}} l*{3}{c}} \hline\hline
Dependent Variable: & \multicolumn{2}{c}{Standardized Sentiment} & Number of Words \\ \cline{2-3} \cline{4-4}
                    &\multicolumn{1}{c}{(1)}   &\multicolumn{1}{c}{(2)}   &\multicolumn{1}{c}{(3)}   \\
\hline
Number of Words     &    -0.00033***&    -0.00051***&               \\
                    &  (0.000043)   &  (0.000081)   &               \\
New Forum Member    &       0.033***&       0.059** &       -1.77   \\
                    &     (0.012)   &     (0.023)   &      (1.45)   \\
Experienced Forum Member&       0.038***&      -0.033*  &        1.43   \\
                    &     (0.012)   &     (0.020)   &      (1.46)   \\
Agora Vendor        &      -0.063***&        0.26***&       -2.16   \\
                    &     (0.017)   &     (0.097)   &      (5.40)   \\
First Post in Thread&       -0.10***&      -0.057***&        38.8***\\
                    &     (0.017)   &     (0.021)   &      (2.97)   \\
Positive Karma      &      0.0011***&     0.00036   &       0.084   \\
                    &   (0.00012)   &   (0.00028)   &     (0.068)   \\
Negative Karma      &     -0.0025***&    -0.00045   &      -0.063   \\
                    &   (0.00021)   &   (0.00038)   &     (0.065)   \\
Missing Karma Data  &      -0.070*  &       0.033   &       -6.99** \\
                    &     (0.040)   &     (0.051)   &      (2.78)   \\
Product Categories Subforum&        0.20***&        0.13***&        12.6***\\
                    &     (0.013)   &     (0.030)   &      (2.49)   \\
Product Offers Subforum&        0.23***&       0.062***&       -6.83***\\
                    &     (0.016)   &     (0.023)   &      (1.54)   \\
Vendor Discussion Subforum&        0.11***&       0.065***&        5.16***\\
                    &     (0.011)   &     (0.017)   &      (1.02)   \\
Constant            &       -0.10   &        0.15   &       -0.94   \\
                    &     (0.089)   &      (0.23)   &      (6.63)   \\
\hline
Observations        &       61660   &       61660   &       61660   \\
R-Squared           &        0.01   &        0.07   &        0.08   \\
Fixed Effects       &        Time   &Time+Author+Vendor   &Time+Author+Vendor   \\
Mean of Dependent Variable&        0.00   &        0.00   &       41.56   \\
SD of Dependent Variable&        0.98   &        0.98   &       68.45   \\
\hline\hline
\multicolumn{4}{l}{\footnotesize Standard errors in parentheses}\\
\multicolumn{4}{l}{\footnotesize * p<0.10, ** p<0.05, *** p<0.01}\\
\end{tabular}
}

%% file: baselineDemandRegs_fe.tex
\begin{tabular}{@{\extracolsep{4pt}} l c c c c c@{}} 

 \hline \multicolumn{6}{l}{Dependent Variable: Number of Sales per Week} \\ \hline
& (1) & (2) & (3) & (4) & (5)  \\
\hline Average Item Rating & 1.6170*** & 1.6179*** & 1.6045*** & 1.5942*** & 1.5549***\\ 
 & (0.0795) & (0.0799) & (0.0789) & (0.0736) & (0.0671)\\ 
Avg. Vendor Review Sentiment & 1.2353*** & 1.2317*** & 1.2107*** & 1.0430 & 1.0875***\\ 
 & (0.0410) & (0.0404) & (0.0381) & (0.0343) & (0.0351)\\ 
log(Number of Vendor Reviews) & 1.0817*** & 1.0395*** & 1.0283** & 0.7695*** & 0.8447***\\ 
 & (0.0095) & (0.0112) & (0.0111) & (0.0084) & (0.0104)\\ 
Avg. Vendor Forum Sentiment &  & 1.1254*** & 1.1191*** & 1.1067*** & 1.1074***\\ 
 &  & (0.0327) & (0.0323) & (0.0317) & (0.0308)\\ 
log(Number of Forum Mentions) &  & 1.1001*** & 1.0779*** & 0.9780 & 0.9993\\ 
 &  & (0.0197) & (0.0201) & (0.0187) & (0.0196)\\ 
log(Price) (BTC) &  &  & 1.0790*** & 1.0010 & 0.9942\\ 
 &  &  & (0.0230) & (0.0142) & (0.0148)\\ 
$\mathbbm{1}\{\textrm{No Finalize Early on Product}\}$ &  &  & 1.4923*** & 1.3081*** & 1.2730***\\ 
 &  &  & (0.0637) & (0.0549) & (0.0552)\\ 
Vendor Rating &  &  &  & 1.0227 & 0.9899\\ 
 &  &  &  & (0.0544) & (0.0494)\\ 
 \hline 
Observations & 328838 & 328838 & 328659 & 328659 & 328659\\ 
Log-likelihood & -234193.7 & -234024.8 & -232795.2 & -220724.9 & -208663.6\\ 
Item Controls & False & False & True & True & True\\ 
Vendor Controls & False & False & False & True & True\\ 
Time FE & False & False & False & False & True\\ 
\hline

\end{tabular} 

%% file: diffEffectsDemandRegs_fe.tex
\begin{tabular}{@{\extracolsep{4pt}} l c c c c@{}} 

 \hline \multicolumn{5}{l}{Dependent Variable: Number of Sales per Week} \\ \hline
& (1) & (2) & (3) & (4)  \\
\hline Avg. Vendor Review Sentiment & 0.9866 &  &  & \\ 
 & (0.0406) &  &  & \\ 
Avg. Vendor Forum Sentiment & 1.0222 &  &  & 1.1064***\\ 
 & (0.0282) &  &  & (0.0307)\\ 
(Avg. Vendor Review Sentiment) $\times$ log(Number of Vendor Reviews) & 1.0553*** &  &  & \\ 
 & (0.0193) &  &  & \\ 
(Avg. Vendor Forum Sentiment) $\times$ log(Number of Forum Mentions) & 1.0915*** &  &  & \\ 
 & (0.0250) &  &  & \\ 
Avg. Vendor Review Sentiment of Inexperienced Customers &  & 1.1036*** &  & \\ 
 &  & (0.0316) &  & \\ 
Avg. Vendor Review Sentiment of Experienced Customers &  & 0.9940 &  & \\ 
 &  & (0.0328) &  & \\ 
Avg. Sentiment of Experienced Forum Users &  & 1.0909*** &  & \\ 
 &  & (0.0312) &  & \\ 
Avg. Sentiment of Inexperienced Forum Users &  & 1.0986*** &  & \\ 
 &  & (0.0302) &  & \\ 
$\mathbbm{1}\{\textrm{Finalize Early}\}*(\textrm{Vendor Review Sentiment})$ &  &  & 1.1435** & \\ 
 &  &  & (0.0617) & \\ 
$\mathbbm{1}\{\textrm{No Finalize Early}\}*(\textrm{Vendor Review Sentiment})$ &  &  & 1.0600 & \\ 
 &  &  & (0.0408) & \\ 
$\mathbbm{1}\{\textrm{Finalize Early}\}*(\textrm{Vendor Forum Sentiment})$ &  &  & 1.2193*** & \\ 
 &  &  & (0.0694) & \\ 
$\mathbbm{1}\{\textrm{No Finalize Early}\}*(\textrm{Vendor Forum Sentiment})$ &  &  & 1.0861*** & \\ 
 &  &  & (0.0324) & \\ 
Avg. Product Sentiment &  &  &  & 1.0262\\ 
 &  &  &  & (0.0193)\\ 
Avg. Sentiment of Other Products listed by Vendor &  &  &  & 1.1011***\\ 
 &  &  &  & (0.0376)\\ 
 \hline 
Observations & 328659 & 328659 & 328659 & 328659\\ 
Log-likelihood & -208583.0 & -207864.0 & -208651.9 & -208652.3\\ 
P-value of Equality of Review Sentiments & N/A & 0.0241 & 0.2345 & 0.0802\\ 
P-value of Equality of Forum Sentiments & N/A & 0.8679 & 0.0581 & N/A\\ 
Item Controls & True & True & True & True\\ 
Vendor Controls & True & True & True & True\\ 
Time FE & True & True & True & True\\ 
\hline

\end{tabular} 

%% file: VendorListingRegs.tex
\begin{tabular}{@{\extracolsep{4pt}} l c c c c@{}} 

 \hline & (1) & (2) & (3) & (4)  \\
 & \multicolumn{2}{c}{\# of Listings (GRAMS)} & \multicolumn{2}{c}{\# of Listings (Agora Scrapes)} \\ \cline{2-3} \cline{4-5}
 Avg. Vendor Review Sentiment & 0.9920 & 0.9909 & 0.9189* & 0.9166*\\ 
 & (0.0164) & (0.0170) & (0.0434) & (0.0431)\\ 
log(Number of Vendor Reviews) & 1.0183*** & 1.0187*** & 1.5613*** & 1.5611***\\ 
 & (0.0039) & (0.0041) & (0.0236) & (0.0236)\\ 
Avg. Vendor Forum Sentiment & 1.0475 & 1.0349 & 1.0107 & 1.0037\\ 
 & (0.0933) & (0.1027) & (0.0390) & (0.0473)\\ 
log(Number of Forum Mentions) & 1.0045 & 1.0034 & 0.9864 & 0.9853\\ 
 & (0.0105) & (0.0121) & (0.0242) & (0.0258)\\ 
\% No Finalize Early Listings & 0.9395 & 0.9335 & 0.7590*** & 0.7537***\\ 
 & (0.0405) & (0.0451) & (0.0484) & (0.0475)\\ 
Vendor Rating & 0.9608 & 0.9625 & 1.0393 & 1.0402\\ 
 & (0.0462) & (0.0488) & (0.0276) & (0.0273)\\ 
 \hline 
 $\alpha$ (Dispersion Parameter) &  & 1.0372 &  & 0.8931\\ 
 &  & (3.4874) &  & (2.9004)\\ 
 \hline 
 Observations & 32443 & 32577 & 42300 & 42456\\ 
Log-likelihood & -88932.5 & -98952.5 & -127234.7 & -138740.4\\ 
Vendor Controls & True & True & True & True\\ 
Time FE & True & True & True & True\\ 
Vendor-specific Effects & FE & Gamma RE & FE & Gamma RE\\ 
\hline

\end{tabular} 

%% file: baselineDemandRegs_re.tex
\begin{tabular}{@{\extracolsep{4pt}} l c c c c c@{}} 

 \hline \multicolumn{6}{l}{Dependent Variable: Number of Sales per Week} \\ \hline
& (1) & (2) & (3) & (4) & (5)  \\
\hline Average Item Rating & 1.5652*** & 1.5546*** & 1.5267*** & 1.4789*** & 1.4469***\\ 
 & (0.0703) & (0.0700) & (0.0673) & (0.0621) & (0.0562)\\ 
Avg. Vendor Review Sentiment & 1.2589*** & 1.2638*** & 1.2570*** & 1.0592** & 1.0932***\\ 
 & (0.0371) & (0.0370) & (0.0367) & (0.0307) & (0.0297)\\ 
log(Number of Vendor Reviews) & 1.0505*** & 1.0004 & 1.0140 & 0.7311*** & 0.7836***\\ 
 & (0.0103) & (0.0115) & (0.0101) & (0.0100) & (0.0123)\\ 
Avg. Vendor Forum Sentiment &  & 1.2280*** & 1.1671*** & 1.1325*** & 1.1239***\\ 
 &  & (0.0374) & (0.0287) & (0.0263) & (0.0250)\\ 
log(Number of Forum Mentions) &  & 1.1243*** & 1.0839*** & 0.9914 & 0.9763**\\ 
 &  & (0.0140) & (0.0131) & (0.0125) & (0.0119)\\ 
log(Price) (BTC) &  &  & 0.9460*** & 0.9028*** & 0.9038***\\ 
 &  &  & (0.0193) & (0.0146) & (0.0138)\\ 
$\mathbbm{1}\{\textrm{No Finalize Early on Product}\}$ &  &  & 1.5893*** & 1.4309*** & 1.4083***\\ 
 &  &  & (0.0501) & (0.0442) & (0.0447)\\ 
Vendor Rating &  &  &  & 1.0065 & 0.9707\\ 
 &  &  &  & (0.0424) & (0.0349)\\ 
 \hline 
$\alpha$ (Dispersion Parameter) & 4.2601*** & 4.1346*** & 2.8306*** & 2.5414*** & 2.4376***\\ 
 & (1.0787) & (1.0705) & (0.8585) & (0.7883) & (0.7771)\\ 
 \hline 
Observations & 609476 & 609476 & 608072 & 608072 & 608072\\ 
Log-likelihood & -314631.6 & -314184.1 & -309573.0 & -296298.7 & -283914.8\\ 
Item Controls & False & False & True & True & True\\ 
Vendor Controls & False & False & False & True & True\\ 
Time FE & False & False & False & False & True\\ 
\hline

\end{tabular} 

%% file: diffEffectsDemandRegs_re.tex
\begin{tabular}{@{\extracolsep{4pt}} l c c c c@{}} 

 \hline \multicolumn{5}{l}{Dependent Variable: Number of Sales per Week} \\ \hline
& (1) & (2) & (3) & (4)  \\
\hline Avg. Vendor Review Sentiment & 0.9741 &  &  & \\ 
 & (0.0356) &  &  & \\ 
Avg. Vendor Forum Sentiment & 1.0212 &  &  & 1.1227***\\ 
 & (0.0244) &  &  & (0.0249)\\ 
(Avg. Vendor Review Sentiment) $\times$ log(Number of Vendor Reviews) & 1.0603*** &  &  & \\ 
 & (0.0159) &  &  & \\ 
(Avg. Vendor Forum Sentiment) $\times$ log(Number of Forum Mentions) & 1.0945*** &  &  & \\ 
 & (0.0189) &  &  & \\ 
Avg. Vendor Review Sentiment of Inexperienced Customers &  & 1.1403*** &  & \\ 
 &  & (0.0302) &  & \\ 
Avg. Vendor Review Sentiment of Experienced Customers &  & 0.9716 &  & \\ 
 &  & (0.0293) &  & \\ 
Avg. Sentiment of Experienced Forum Users &  & 1.1132*** &  & \\ 
 &  & (0.0259) &  & \\ 
Avg. Sentiment of Inexperienced Forum Users &  & 1.1069*** &  & \\ 
 &  & (0.0246) &  & \\ 
$\mathbbm{1}\{\textrm{Finalize Early}\}*(\textrm{Vendor Review Sentiment})$ &  &  & 1.1421*** & \\ 
 &  &  & (0.0527) & \\ 
$\mathbbm{1}\{\textrm{No Finalize Early}\}*(\textrm{Vendor Review Sentiment})$ &  &  & 1.0667** & \\ 
 &  &  & (0.0343) & \\ 
$\mathbbm{1}\{\textrm{Finalize Early}\}*(\textrm{Vendor Forum Sentiment})$ &  &  & 1.2571*** & \\ 
 &  &  & (0.0527) & \\ 
$\mathbbm{1}\{\textrm{No Finalize Early}\}*(\textrm{Vendor Forum Sentiment})$ &  &  & 1.0919*** & \\ 
 &  &  & (0.0274) & \\ 
Avg. Product Sentiment &  &  &  & 1.0111\\ 
 &  &  &  & (0.0177)\\ 
Avg. Sentiment of Other Products listed by Vendor &  &  &  & 1.1137***\\ 
 &  &  &  & (0.0313)\\ 
 \hline 
$\alpha$ (Dispersion Parameter) & 2.4257*** & 2.7893*** & 2.4359*** & 2.4366***\\ 
 & (0.7731) & (0.8901) & (0.7766) & (0.7758)\\ 
 \hline 
Observations & 608072 & 608072 & 608072 & 608072\\ 
Log-likelihood & -283797.4 & -284679.5 & -283894.5 & -283901.0\\ 
P-value of Equality of Review Sentiments & N/A & 0.0003 & 0.2049 & 0.0054\\ 
P-value of Equality of Forum Sentiments & N/A & 0.8722 & 0.0032 & N/A\\ 
Item Controls & True & True & True & True\\ 
Vendor Controls & True & True & True & True\\ 
Time FE & True & True & True & True\\ 
\hline

\end{tabular} 

%% file: baselineDemandRegs_fe_allposts.tex
\begin{tabular}{@{\extracolsep{4pt}} l c c c c c@{}} 

 \hline \multicolumn{6}{l}{Dependent Variable: Number of Sales per Week} \\ \hline
& (1) & (2) & (3) & (4) & (5)  \\
\hline Average Item Rating & 1.6170*** & 1.6203*** & 1.6066*** & 1.5965*** & 1.5561***\\ 
 & (0.0795) & (0.0797) & (0.0787) & (0.0737) & (0.0673)\\ 
Avg. Vendor Review Sentiment & 1.2353*** & 1.2270*** & 1.2073*** & 1.0446 & 1.0854**\\ 
 & (0.0410) & (0.0405) & (0.0381) & (0.0344) & (0.0351)\\ 
log(Number of Vendor Reviews) & 1.0817*** & 1.0440*** & 1.0313*** & 0.7762*** & 0.8517***\\ 
 & (0.0095) & (0.0117) & (0.0114) & (0.0083) & (0.0102)\\ 
Avg. Vendor Forum Sentiment &  & 1.0704** & 1.0677** & 1.0335 & 1.0652**\\ 
 &  & (0.0314) & (0.0315) & (0.0303) & (0.0309)\\ 
log(Number of Forum Mentions) &  & 1.0785*** & 1.0614*** & 0.9581*** & 0.9676**\\ 
 &  & (0.0158) & (0.0158) & (0.0148) & (0.0153)\\ 
log(Price) (BTC) &  &  & 1.0787*** & 1.0015 & 0.9943\\ 
 &  &  & (0.0228) & (0.0143) & (0.0149)\\ 
$\mathbbm{1}\{\textrm{No Finalize Early on Product}\}$ &  &  & 1.4951*** & 1.3125*** & 1.2770***\\ 
 &  &  & (0.0636) & (0.0548) & (0.0551)\\ 
Vendor Rating &  &  &  & 1.0259 & 0.9911\\ 
 &  &  &  & (0.0548) & (0.0496)\\ 
 \hline 
Observations & 328838 & 328838 & 328659 & 328659 & 328659\\ 
Log-likelihood & -234193.7 & -234074.1 & -232830.9 & -220715.8 & -208650.4\\ 
Item Controls & False & False & True & True & True\\ 
Vendor Controls & False & False & False & True & True\\ 
Time FE & False & False & False & False & True\\ 
\hline

\end{tabular} 

%% file: baselineDemandRegs_fe_linear.tex
\begin{tabular}{@{\extracolsep{4pt}} l c c c c c@{}} 

 \hline \multicolumn{6}{l}{Dependent Variable: Number of Sales per Week} \\ \hline
& (1) & (2) & (3) & (4) & (5)  \\
\hline Average Item Rating & 0.09004*** & 0.08963*** & 0.08746*** & 0.08712*** & 0.07676***\\ 
 & (0.008049) & (0.008078) & (0.008088) & (0.007783) & (0.007813)\\ 
Avg. Vendor Review Sentiment & 0.05481*** & 0.05472*** & 0.05680*** & 0.04302*** & 0.04037***\\ 
 & (0.006844) & (0.006823) & (0.006907) & (0.007065) & (0.006911)\\ 
log(Number of Vendor Reviews) & 0.02588*** & 0.01036*** & 0.007108** & -0.1367*** & -0.08478***\\ 
 & (0.003117) & (0.003395) & (0.003430) & (0.004633) & (0.004636)\\ 
Avg. Vendor Forum Sentiment &  & 0.04773*** & 0.04679*** & 0.06529*** & 0.06433***\\ 
 &  & (0.008898) & (0.008791) & (0.009001) & (0.008862)\\ 
log(Number of Forum Mentions) &  & 0.04591*** & 0.03938*** & 0.01029 & 0.007857\\ 
 &  & (0.006889) & (0.006919) & (0.006665) & (0.006563)\\ 
log(Price) (BTC) &  &  & 0.01671*** & -0.0005493 & -0.0007265\\ 
 &  &  & (0.002963) & (0.002862) & (0.002915)\\ 
$\mathbbm{1}\{\textrm{No Finalize Early on Product}\}$ &  &  & 0.1386*** & 0.06953*** & 0.07066***\\ 
 &  &  & (0.01493) & (0.01499) & (0.01464)\\ 
Vendor Rating &  &  &  & -0.02178*** & -0.03958***\\ 
 &  &  &  & (0.002777) & (0.003248)\\ 
 \hline 
Observations & 609476 & 609476 & 608072 & 608072 & 608072\\ 
Log-likelihood & -783802.4 & -783654.8 & -781664.4 & -776564.3 & -771370.0\\ 
Item Controls & False & False & True & True & True\\ 
Vendor Controls & False & False & False & True & True\\ 
Time FE & False & False & False & False & True\\ 
\hline

\end{tabular} 

%% file: diffEffectsDemandRegs_fe_linear.tex
\begin{tabular}{@{\extracolsep{4pt}} l c c c c@{}} 

 \hline \multicolumn{5}{l}{Dependent Variable: Number of Sales per Week} \\ \hline
& (1) & (2) & (3) & (4)  \\
\hline Avg. Vendor Review Sentiment & 0.001292 &  &  & \\ 
 & (0.009058) &  &  & \\ 
Avg. Vendor Forum Sentiment & 0.004517 &  &  & 0.06428***\\ 
 & (0.009232) &  &  & (0.008843)\\ 
(Avg. Vendor Review Sentiment) $\times$ log(Number of Vendor Reviews) & 0.02677*** &  &  & \\ 
 & (0.005717) &  &  & \\ 
(Avg. Vendor Forum Sentiment) $\times$ log(Number of Forum Mentions) & 0.07404*** &  &  & \\ 
 & (0.01079) &  &  & \\ 
Avg. Vendor Review Sentiment of Inexperienced Customers &  & 0.03640*** &  & \\ 
 &  & (0.006106) &  & \\ 
Avg. Vendor Review Sentiment of Experienced Customers &  & 0.01097 &  & \\ 
 &  & (0.007232) &  & \\ 
Avg. Sentiment of Experienced Forum Users &  & 0.06989*** &  & \\ 
 &  & (0.01217) &  & \\ 
Avg. Sentiment of Inexperienced Forum Users &  & 0.05491*** &  & \\ 
 &  & (0.009371) &  & \\ 
$\mathbbm{1}\{\textrm{Finalize Early}\}*(\textrm{Vendor Review Sentiment})$ &  &  & 0.02957*** & \\ 
 &  &  & (0.008432) & \\ 
$\mathbbm{1}\{\textrm{No Finalize Early}\}*(\textrm{Vendor Review Sentiment})$ &  &  & 0.05058*** & \\ 
 &  &  & (0.01018) & \\ 
$\mathbbm{1}\{\textrm{Finalize Early}\}*(\textrm{Vendor Forum Sentiment})$ &  &  & 0.09639*** & \\ 
 &  &  & (0.01191) & \\ 
$\mathbbm{1}\{\textrm{No Finalize Early}\}*(\textrm{Vendor Forum Sentiment})$ &  &  & 0.04907*** & \\ 
 &  &  & (0.01143) & \\ 
Avg. Product Sentiment &  &  &  & 0.02109***\\ 
 &  &  &  & (0.007268)\\ 
Avg. Sentiment of Other Products listed by Vendor &  &  &  & 0.03476***\\ 
 &  &  &  & (0.007398)\\ 
 \hline 
Observations & 608072 & 608072 & 608072 & 608072\\ 
Log-likelihood & -771203.9 & -771701.9 & -771360.1 & -771365.6\\ 
P-value of Equality of Review Sentiments & N/A & 0.0124 & 0.0973 & 0.1933\\ 
P-value of Equality of Forum Sentiments & N/A & 0.3483 & 0.0034 & N/A\\ 
Item Controls & True & True & True & True\\ 
Vendor Controls & True & True & True & True\\ 
Time FE & True & True & True & True\\ 
\hline

\end{tabular} 